\documentclass{article}

\usepackage[utf8]{inputenc}
\usepackage{lipsum}
\usepackage{forloop}
\usepackage{caption}
\usepackage{multirow}
\usepackage{soul}

\usepackage[margin=20mm]{geometry}
\usepackage{multicol}
\usepackage{titlesec}
\renewcommand{\thesection}{\Roman{section}}
\renewcommand{\thesubsection}{\thesection}
\renewcommand{\thesubsubsection}{\thesubsection-\arabic{subsubsection}}
\titleformat{\section}{\bf\normalsize\filcenter}{\thesection.}{1ex}{}
\titleformat{\subsection}{\bf\normalsize}{}{0ex}{}
\titleformat{\subsubsection}{\bf\normalsize}{\thesubsubsection:}{1ex}{}
\newcommand{\subsubsubsection}[1]{\paragraph{#1}\hspace{-1ex}:\hspace{1ex}}

\newcommand{\breakl}{\par\noindent\rule[1ex]{0.5\textwidth}{0.4pt}}

\newcommand{\breakc}{\begin{center}\par\noindent\rule[1ex]{0.75\textwidth}{0.4pt}\end{center}}

\usepackage{amsmath}
\usepackage{siunitx}
\DeclareSIUnit\angstrom{\text {Å}}
\newcommand{\angstrom}{\si{\angstrom}}%\angstrom needs to be in math mode if used in a paragraph to avoid spacing issues

\newcommand{\n}{\nonumber\\ }
\newcommand{\nn}[1][3]{\nonumber\\&\hspace{#1ex}}
\newcommand{\mat}[1]{\begin{matrix}#1\end{matrix}}

\usepackage[hidelinks]{hyperref}
\usepackage{cite}
\usepackage{tikz}
\usetikzlibrary{shapes}

\tikzstyle{wire}=[thick]
\tikzstyle{gate}=[rectangle, minimum width=0.8cm, minimum height=0.8cm, text width=0cm, inner sep=0pt, draw=black, fill=white]
\tikzstyle{state}=[rectangle, minimum width=0.8cm, minimum height=0.8cm, text width=0cm, inner sep=0pt, draw=none, fill=white]
\tikzstyle{target}=[circle, minimum width=0.5cm, minimum height=0.5cm, text width=0cm, inner sep=0pt, draw=black, fill=white]
\tikzstyle{control}=[circle, minimum width=0.2cm, minimum height=0.2cm, text width=0cm, inner sep=0pt, draw=black, fill=black]
\tikzstyle{controlzero}=[circle, minimum width=0.2cm, minimum height=0.2cm, text width=0cm, inner sep=0pt, draw=black, fill=white]
\tikzstyle{measure}=[diamond, minimum width=0.9cm, minimum height=0.9cm, text width=0cm, inner sep=0pt, draw=black, fill=white]

\newcommand{\gate}[3][]{
\node at (#3,-#2)(gate)[gate]{};
\node at (#3,-#2){$#1$};
}
\newcommand{\state}[3][|0\rangle]{
\node at (#3,-#2)(state)[state]{};
\node at (#3,-#2){$#1$};
}
\newcommand{\target}[2]{
\node at (#2,-#1)(target)[target]{};
\draw[wire](target.north)--(target.south);
\draw[wire](target.east)--(target.west);
}
\newcommand{\control}[3][control]{
\node at (#3,-#2)(#1)[control]{};
}
\newcommand{\controlzero}[2]{
\node at (#2,-#1)(control)[controlzero]{};
}
\newcommand{\cnot}[3]{
\control{#1}{#3}
\target{#2}{#3}
\draw[wire](control)--(target);
}
\newcommand{\cgate}[4][]{
\control{#2}{#4}
\gate[#1]{#3}{#4}
\draw[wire](control)--(gate);
}
\newcommand{\bundle}[2]{
\node at (#2,-#1){/};
}
\newcommand{\cz}[3]{
\control{#1}{#3}
\control[control2]{#2}{#3}
\draw[wire](control)--(control2);
}
\newcommand{\measure}[3][?]{
\node at (#3,-#2)(measure)[measure]{};
\node at (#3,-#2){$#1$};
}
\newcommand{\qubit}[4][]{
\draw[wire,#1](#3,-#2)--(#4,-#2);
}

\newcounter{qubitnum}
\newenvironment{circuit}[2]{
\begin{tikzpicture}
\forloop{qubitnum}{0}{\value{qubitnum} < #1}{\qubit{\arabic{qubitnum}}{0}{#2}}
}{
\end{tikzpicture}
}

\renewcommand{\title}[1]{
\begin{center}
{\large\bf{#1}}
\end{center}
}
\newcommand{\authors}[2]{
\begin{center}
#1\\\vspace{1ex}
{\it {\footnotesize#2}}
\end{center}
}
\renewcommand{\abstract}[1]{{\bf{#1}}}

\newcommand{\references}[1]{
\breakc
\begin{multicols}{2}
\let\oldsection\section
\renewcommand{\section}[2]{} %\bibligraphy calls \section so use this to prevent including a title
{\scriptsize\bibliography{#1}}
\let\section\oldsection
\end{multicols}
}

\newcommand{\supinf}[1][Supplementary Information]{
\newpage
\appendix
\titleformat{\section}{\bf\large\filcenter}{}{0ex}{}
\section{#1}
\renewcommand{\thesection}{SI - \Roman{section}}
\titleformat{\section}{\bf\normalsize\filcenter}{\thesection.}{1ex}{}
\setcounter{section}{0}
\renewcommand{\theequation}{SI.\arabic{equation}}
\setcounter{equation}{0}
\renewcommand{\thefigure}{SI.\arabic{figure}}
\setcounter{figure}{0}
\renewcommand{\thetable}{SI.\arabic{table}}
\setcounter{table}{0}
\renewcommand{\thesubsection}{\thesection \alph{subsection}}
\titleformat{\subsection}{\bf\normalsize}{\thesubsection:}{1ex}{}
}

\newcommand{\includefigflow}[1]{
    \begin{figure*}[#1]
    \centering
    \includegraphics[width=\linewidth]{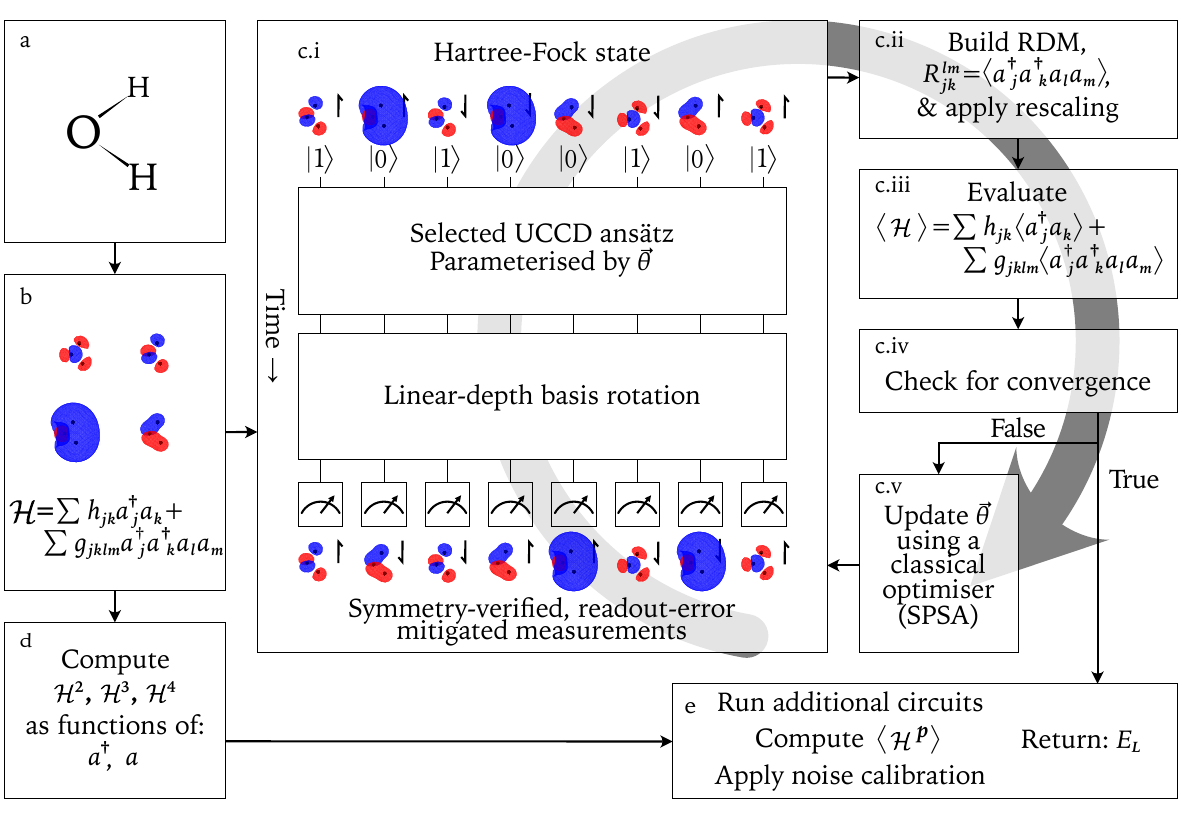}
    \caption{Flow chart of the overall approach employed in this work.
    a) Definition of the molecular system.
    b) Calculation of the molecular orbitals and electronic Hamiltonian.
    c) Hybrid minimisation loop to approximate the ground-state on the quantum device.
    d) Powers of the Hamiltonian can be evaluated independently of (i.e. in parallel with) the minimisation loop.
    e) Calculation of the Hamiltonian moments, $\langle\mathcal{H}\rangle$ and hence the Lanczos cluster expansion corrected energy estimate, $E_L$.
    For additional detail on the above steps see Section~\ref{sec:methods} and the supplementary information.}
    \label{fig:flow}
    \end{figure*}
}
\newcommand{\includefigcircuit}[1]{
    \begin{figure*}[#1]
    \centering
    \includegraphics[width=0.95\linewidth]{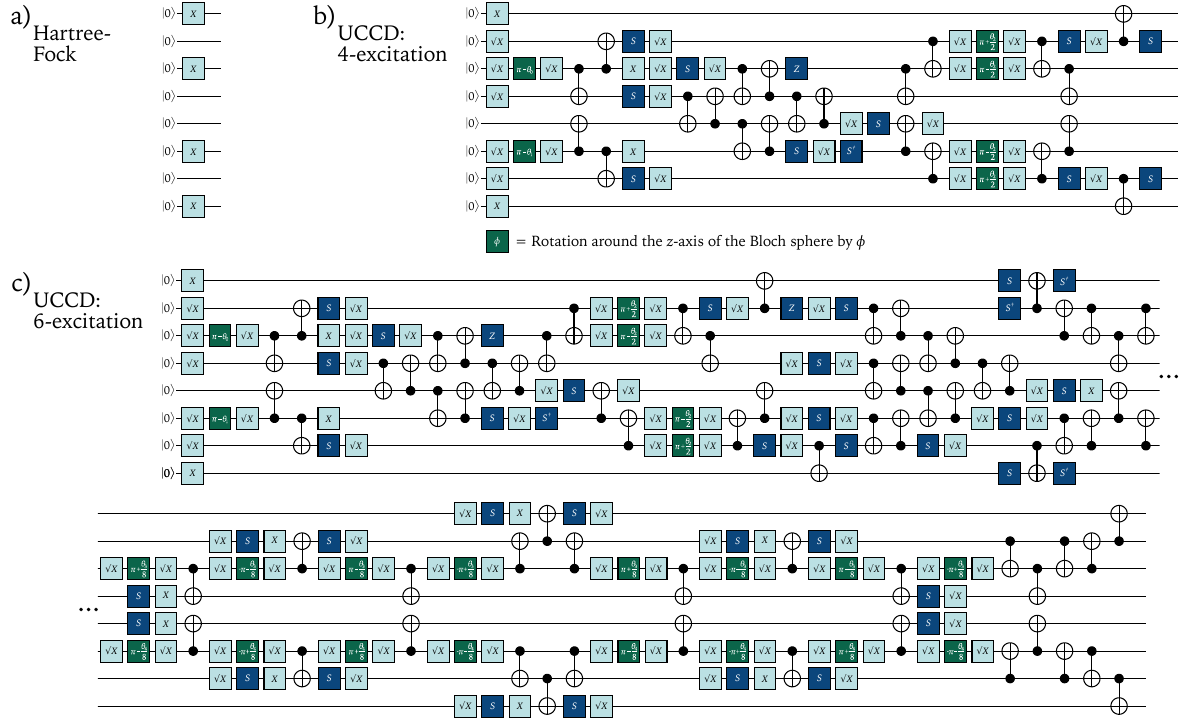}
    \caption{
    Trial-state circuits used in this work: a) Minimal circuit to prepare the Hartree-Fock state. b) 4-excitation trial-state circuit with depth $D=25$. Setting all $\theta_j=0$ produces the Hartree-Fock state for reference calibration. c) 6-excitation trial-state circuit with depth $D=74$. These circuits are followed by linear depth basis rotations (see \ref{app:RDM}), then computational basis measurements.
    }
    \label{fig:circuits}
    \end{figure*}
}
\newcommand{\includefigres}[1]{
    \begin{figure*}[#1]
    \centering
    \includegraphics[width=0.95\linewidth]{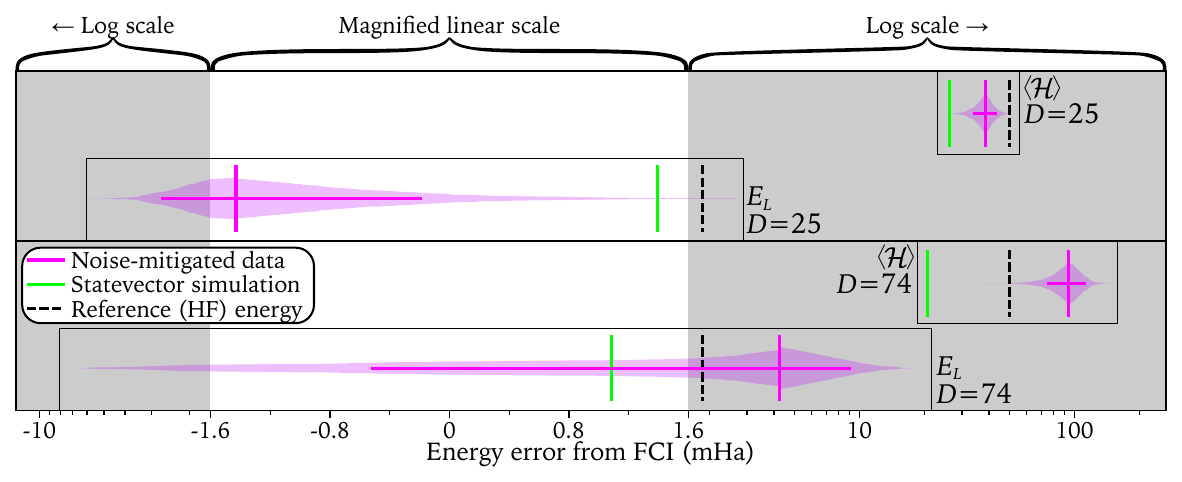}
    \caption{
    Energy estimation results for the water molecule based on measuring $\langle\mathcal{H}\rangle$ and $E_L$ for the 4-excitation ($D=25$) and 6-excitation ($D=74$) circuits presented in Figure~\ref{fig:circuits}b and c.
    The noiseless simulation results are shown for the trial-state (in green) and the reference-state (in dashed black).
    Noise-mitigated results, taken from the quantum device \emph{ibmq\_kolkata}, are marked by vertical magenta bars.
    The horizontal magenta lines are one standard deviation of the 500 statistical bootstrapping results which are shown as a cumulative distribution (i.e. the height of the distribution is proportional to the number of results that are towards to the tail of the distribution).
    Note that the horizontal axis uses a linear scale in the target accuracy window $[-1.6,1.6]$ mHa and a logarithmic scale outside this window.
    For comparison, the $[E_L,D=25]$ results from the quantum hardware are at $-1.43$ mHa while the statevector and reference values are $1.39$ and $1.85$ mHa respectively.
    Since $E_L$ is not an upper-bound, negative deviations are not unexpected.
    }
    \label{fig:results}
    \end{figure*}
}
\newcommand{\includefigfswap}[2][1]{
    \begin{figure*}[#2]
    \centering
    a.
    \begin{minipage}[t]{0.9\linewidth}
    $$
    G_{jklm}(\theta)=
    \scalebox{#1}{\begin{minipage}[c][6.5cm]{7.5cm}\centering
    \begin{tikzpicture}
    \qubit[red]{0}{0}{6}\state[q_j]{0}{0}
    \qubit{1}{0}{6}\bundle{1}{0.5}
    \qubit[green]{2}{0}{6}\state[q_k]{2}{0}
    \qubit{3}{0}{6}\bundle{3}{0.5}
    \qubit[blue]{4}{0}{6}\state[q_l]{4}{0}
    \qubit{5}{0}{6}\bundle{5}{0.5}
    \qubit[orange]{6}{0}{6}\state[q_m]{6}{0}
    \cz{0}{1}{1}\cz{5}{6}{1}
    \cnot{0}{6}{2}\target{2}{2}\target{4}{2}
    \cgate[\theta]{6}{0}{3}\control{2}{3}\controlzero{4}{3}
    \cnot{0}{6}{4}\target{2}{4}\target{4}{4}
    \cz{0}{1}{5}\cz{5}{6}{5}
    \state[q_j]{0}{6}
    \bundle{1}{5.5}
    \state[q_k]{2}{6}
    \bundle{3}{5.5}
    \state[q_l]{4}{6}
    \bundle{5}{5.5}
    \state[q_m]{6}{6}
    \end{tikzpicture}
    \end{minipage}}
    \longrightarrow
    \scalebox{#1}{\begin{minipage}[c][6.5cm]{7.5cm}\centering
    \begin{tikzpicture}
    \draw[wire,red](0,0)--(0.7,0)--(1.3,-1)--(6,-1);
    \draw[wire](0,-1)--(0.7,-1)--(1.3,0)--(6,0);
    \draw[wire,green](0,-2)--(6,-2);
    \draw[wire](0,-3)--(0.7,-3)--(1.3,-4)--(1.7,-4)--(2.3,-5)--(6,-5);
    \draw[wire,blue](0,-4)--(0.7,-4)--(1.3,-3)--(6,-3);
    \draw[wire](0,-5)--(0.7,-5)--(1.3,-6)--(6,-6);
    \draw[wire,orange](0,-6)--(0.7,-6)--(1.3,-5)--(1.7,-5)--(2.3,-4)--(6,-4);
    \state[q_j]{0}{0}
    \bundle{1}{0.5}
    \state[q_k]{2}{0}
    \bundle{3}{0.5}
    \state[q_l]{4}{0}
    \bundle{5}{0.5}
    \state[q_m]{6}{0}
    \node at (1,-0.5)[circle, minimum width=0.6cm, minimum height=0.6cm, text width=0cm, inner sep=0pt, draw=black, fill=white]{};\node at (1,-0.5){$f$};
    \node at (1,-3.5)[circle, minimum width=0.6cm, minimum height=0.6cm, text width=0cm, inner sep=0pt, draw=black, fill=white]{};\node at (1,-3.5){$f$};
    \node at (1,-5.5)[circle, minimum width=0.6cm, minimum height=0.6cm, text width=0cm, inner sep=0pt, draw=black, fill=white]{};\node at (1,-5.5){$f$};
    \node at (2,-4.5)[circle, minimum width=0.6cm, minimum height=0.6cm, text width=0cm, inner sep=0pt, draw=black, fill=white]{};\node at (2,-4.5){$f$};
    \cnot{1}{4}{3}\target{2}{3}\target{3}{3}
    \cgate[\theta]{4}{1}{4}\control{2}{4}\controlzero{3}{4}
    \cnot{1}{4}{5}\target{2}{5}\target{3}{5}
    \state[q_j]{1}{6}
    \bundle{0}{5.5}
    \state[q_k]{2}{6}
    \bundle{5}{5.5}
    \state[q_l]{3}{6}
    \bundle{6}{5.5}
    \state[q_m]{4}{6}
    \end{tikzpicture}
    \end{minipage}}
    $$
    \end{minipage}
    
    \vspace{1cm}b.
    \begin{minipage}[t]{0.9\linewidth}
    $$
    \scalebox{#1}{\begin{minipage}[c]{3.5cm}\centering
    \begin{tikzpicture}
    \draw[wire,red](0,0)--(0.7,0)--(1.3,-1)--(2,-1);
    \draw[wire](0,-1)--(0.7,-1)--(1.3,0)--(2,0);
    \bundle{1}{0.4}
    \node at (1,-0.5)[circle, minimum width=0.6cm, minimum height=0.6cm, text width=0cm, inner sep=0pt, draw=black, fill=white]{};\node at (1,-0.5){$f$};
    \bundle{0}{1.6}
    \end{tikzpicture}
    \end{minipage}}
    =
    \scalebox{#1}{\begin{minipage}[c]{12.5cm}\centering
    \begin{tikzpicture}
    \qubit[red]{0}{0}{2}\qubit[red!50]{0}{2}{3}\qubit{0}{3}{11}
    \qubit{1}{0}{2}\qubit[red!50]{1}{2}{3}\qubit[red]{1}{3}{4}\qubit[red!50]{1}{4}{5}\qubit{1}{5}{11}
    \qubit{2}{0}{4}\qubit[red!50]{2}{4}{5}\qubit[red]{2}{5}{6}\qubit{2}{7}{11}
    \qubit{3}{0}{6}\qubit[red]{3}{7}{8}\qubit[red!50]{3}{8}{9}\qubit{3}{9}{11}
    \qubit{4}{0}{8}\qubit[red!50]{4}{8}{9}\qubit[red]{4}{9}{11}
    \draw[wire,red,dashed](6,-2)--(7,-3);
    \draw[wire,dashed](6,-3)--(7,-2);
    \gate[H]{0}{1}
    \cnot{0}{1}{2}
    \cnot{1}{0}{3}
    \cnot{1}{2}{4}
    \cnot{2}{1}{5}
    \cnot{3}{4}{8}
    \cnot{4}{3}{9}
    \gate[H]{4}{10}
    \end{tikzpicture}
    \end{minipage}}
    $$
    \end{minipage}
    \caption{
    a. Conversion of the non-local fermionic double-excitation to a local operation using fermionic swap operations, $f$, on linear connectivity.
    The $\theta$ gate is a $Y$-rotation by angle $\theta$ and when combined with the CNOT gates, the controlled version can be decomposed to respect the linear connectivity as in Figure~\ref{fig:G4}
    The C$Z$ interactions are between each qubit in the bundle and one of the target qubits.
    The output of the two circuits is identical up to reordering with fermionic swaps.
    b. Decomposition of the fermionic swap gate.
    Note that when many swaps are chained together in this way, the fermionic swap is cheaper to implement than the standard swap in both number of CNOTs and depth.
    }
    \label{fig:fswap}
\end{figure*}
}
\newcommand{\includefigGtwo}[1]{
\begin{figure*}[#1]
    \centering
    \begin{tabular}{ccc|c|c|}
        Operator&Circuit&\multicolumn{1}{c}{}&\multicolumn{1}{c}{Bitstring}&\multicolumn{1}{c}{Eigenvalue}\\\\\cline{4-5}
        \multirow{4}{*}{$\mathrm{Re}(a^\dagger_ja_{j+1})$}&
        \multirow{4}{*}{\scalebox{0.85}{
        \begin{minipage}[c][2.5cm]{6.5cm}\centering
        \begin{circuit}{2}{6}
            \gate[H]{0}{1}
            \cnot{0}{1}{2}
            \gate[\frac{\pi}{4}]{0}{3}
            \gate[\frac{\pi}{4}]{1}{3}
            \cnot{0}{1}{4}
            \gate[H]{0}{5}
            \measure{0}{6}
            \measure{1}{6}
        \end{circuit}
        \end{minipage}}}&
        &00&0\\\cline{4-5}
        &&&01&$1/2$\\\cline{4-5}
        &&&10&$-1/2$\\\cline{4-5}
        &&&11&0\\\cline{4-5}\\\cline{4-5}
        \multirow{4}{*}{$\mathrm{Im}(a^\dagger_ja_{j+1})$}&
        \multirow{4}{*}{\scalebox{0.85}{
        \begin{minipage}[c][2.5cm]{6.5cm}\centering
        \begin{circuit}{2}{6}
            \gate[H]{0}{1}
            \gate[S]{1}{1}
            \cnot{0}{1}{2}
            \gate[\frac{\pi}{4}]{0}{3}
            \gate[\frac{\pi}{4}]{1}{3}
            \cnot{0}{1}{4}
            \gate[H]{0}{5}
            \measure{0}{6}
            \measure{1}{6}
        \end{circuit}
        \end{minipage}}}&
        &00&0\\\cline{4-5}
        &&&01&$-1/2$\\\cline{4-5}
        &&&10&$1/2$\\\cline{4-5}
        &&&11&0\\\cline{4-5}\\\cline{4-5}
        \multirow{2}{*}{$a^\dagger_ja_j$}&
        \multirow{2}{*}{\scalebox{0.85}{
        \begin{minipage}[c][1cm]{2.5cm}\centering
        \begin{circuit}{1}{1}
            \measure{0}{1}
        \end{circuit}
        \end{minipage}}}&
        &0&0\\\cline{4-5}
        &&&1&1\\\cline{4-5}
    \end{tabular}
    \caption{Measurement circuits for single-excitation operators, $a^\dagger_ja_k$. When $k\notin\{j,j\pm1\}$, fermionic swap operations can be used to reorder the logical qubits, then the circuits above can be applied. The $\frac{\pi}{4}$ gates are rotations about the $y$-axis of the Bloch sphere by $\frac{\pi}{4}$.}
    \label{fig:G2}
\end{figure*}
}
\newcommand{\includefigGfour}[2][1]{
\begin{figure*}[#2]
    \begin{align*}
    \scalebox{#1}{\begin{minipage}[c][3.5cm]{4.5cm}\centering
    \begin{circuit}{4}{4}
    \cnot{0}{3}{1}\target{1}{1}\target{2}{1}
    \cgate[\theta]{3}{0}{2}\control{1}{2}\controlzero{2}{2}
    \cnot{0}{3}{3}\target{1}{3}\target{2}{3}
    \end{circuit}
    \end{minipage}}&=
    \scalebox{#1}{\begin{minipage}[c][3.5cm]{17.5cm}\centering
    \begin{circuit}{4}{16}
    \gate[S]{0}{1}
    \gate[S^\dagger]{1}{1}
    \cnot{1}{0}{2}
    \gate[S^\dagger]{0}{3}
    \cnot{2}{1}{3}
    \cnot{1}{2}{4}
    \cnot{2}{3}{5}
    \cnot{1}{2}{6}
    \gate[\frac{\theta}{8}]{2}{7}
    \gate[H]{3}{7}
    \cnot{2}{3}{8}
    \gate[\frac{\theta}{8}]{2}{9}
    \gate[H]{1}{9}
    \cnot{2}{1}{10}
    \gate[-\frac{\theta}{8}]{2}{11}
    \gate[H]{1}{11}
    \cnot{2}{3}{12}
    \gate[-\frac{\theta}{8}]{2}{13}
    \cnot{2}{1}{14}
    \gate[H]{0}{14}
    \cnot{1}{0}{15}
    \state[\scalebox{3}{.\hspace{-1pt}.\hspace{-1pt}.}]{0}{16}
    \state[\scalebox{3}{.\hspace{-1pt}.\hspace{-1pt}.}]{1}{16}
    \state[\scalebox{3}{.\hspace{-1pt}.\hspace{-1pt}.}]{2}{16}
    \state[\scalebox{3}{.\hspace{-1pt}.\hspace{-1pt}.}]{3}{16}
    \end{circuit}
    \end{minipage}
    }
    \\\\&\hspace{10ex}
    \scalebox{#1}{\begin{minipage}[c][3.5cm]{17.5cm}\centering
    \begin{circuit}{4}{14}
    \state[\scalebox{3}{.\hspace{-1pt}.\hspace{-1pt}.}]{0}{0}
    \state[\scalebox{3}{.\hspace{-1pt}.\hspace{-1pt}.}]{1}{0}
    \state[\scalebox{3}{.\hspace{-1pt}.\hspace{-1pt}.}]{2}{0}
    \state[\scalebox{3}{.\hspace{-1pt}.\hspace{-1pt}.}]{3}{0}
    \gate[H]{0}{1}
    \cnot{2}{1}{1}
    \gate[\frac{\theta}{8}]{2}{2}
    \cnot{2}{3}{3}
    \gate[\frac{\theta}{8}]{2}{4}
    \gate[H]{1}{4}
    \cnot{2}{1}{5}
    \gate[-\frac{\theta}{8}]{2}{6}
    \gate[H]{1}{6}
    \cnot{2}{3}{7}
    \gate[-\frac{\theta}{8}]{2}{8}
    \gate[H]{3}{8}
    \cnot{1}{2}{9}
    \cnot{2}{3}{10}
    \cnot{1}{2}{11}
    \cnot{2}{1}{12}
    \cnot{1}{0}{13}
    \end{circuit}
    \end{minipage}
    }
    \end{align*}
    \caption{Decomposition of the local double excitation to linear connectivity. The $\pm\theta/8$ gates are $Y$ rotations by $\pm\theta/8$. This can be improved by prior knowledge of the input state, e.g. correlations between input qubits can be used to remove the long range controls allowing for more efficient decomposition. Including the cost of performing fermionic swaps, the worst-case CNOT count is $4n+3$ for $n$ qubits/spin-orbitals. If it is required that the logical qubits return to their original physical qubit then the cost is $8n-13$ CNOT gates.}
    \label{fig:G4}
\end{figure*}
}
\newcommand{\includefigrouting}[1]{
\begin{figure*}[#1]
    \begin{center}
    \begin{minipage}{0.4\linewidth}
    \centering
    \begin{tikzpicture}
    \draw[wire,red](-0.5,7)--(0.1,7)--(0.9,7)--(1.1,7)--(1.9,7)--(2.1,7)--(2.9,7)--(3.1,7)--(3.9,6)--(4.5,6);
    \draw[wire,red](-0.5,3)--(0.1,3)--(0.9,4)--(1.1,4)--(1.9,5)--(2.1,5)--(2.9,6)--(3.1,6)--(3.9,7)--(4.5,7);
    \draw[wire,green](-0.5,6)--(0.1,6)--(0.9,6)--(1.1,6)--(1.9,6)--(2.1,6)--(2.9,5)--(3.1,5)--(3.9,4)--(4.5,4);
    \draw[wire,green](-0.5,2)--(0.1,2)--(0.9,2)--(1.1,2)--(1.9,3)--(2.1,3)--(2.9,4)--(3.1,4)--(3.9,5)--(4.5,5);
    \draw[wire,blue](-0.5,5)--(0.1,5)--(0.9,5)--(1.1,5)--(1.9,4)--(2.1,4)--(2.9,3)--(3.1,3)--(3.9,2)--(4.5,2);
    \draw[wire,blue](-0.5,1)--(0.1,1)--(0.9,1)--(1.1,1)--(1.9,1)--(2.1,1)--(2.9,2)--(3.1,2)--(3.9,3)--(4.5,3);
    \draw[wire,orange](-0.5,4)--(0.1,4)--(0.9,3)--(1.1,3)--(1.9,2)--(2.1,2)--(2.9,1)--(3.1,1)--(3.9,0)--(4.5,0);
    \draw[wire,orange](-0.5,0)--(0.1,0)--(0.9,0)--(1.1,0)--(1.9,0)--(2.1,0)--(2.9,0)--(3.1,0)--(3.9,1)--(4.5,1);
    \node()[circle,xshift=0.5cm,yshift=3.5cm,fill=white,draw=black,text centered]{$f$};
    \node()[circle,xshift=1.5cm,yshift=4.5cm,fill=white,draw=black,text centered]{$f$};
    \node()[circle,xshift=1.5cm,yshift=2.5cm,fill=white,draw=black,text centered]{$f$};
    \node()[circle,xshift=2.5cm,yshift=1.5cm,fill=white,draw=black,text centered]{$f$};
    \node()[circle,xshift=2.5cm,yshift=3.5cm,fill=white,draw=black,text centered]{$f$};
    \node()[circle,xshift=2.5cm,yshift=5.5cm,fill=white,draw=black,text centered]{$f$};
    \node()[rectangle,xshift=3.5cm,yshift=6.5cm,fill=red!30,draw=black,text centered,minimum height=1.8cm,minimum width=0.8cm]{$U$};
    \node()[rectangle,xshift=3.5cm,yshift=4.5cm,fill=green!30,draw=black,text centered,minimum height=1.8cm,minimum width=0.8cm]{$U$};
    \node()[rectangle,xshift=3.5cm,yshift=2.5cm,fill=blue!30,draw=black,text centered,minimum height=1.8cm,minimum width=0.8cm]{$U$};
    \node()[rectangle,xshift=3.5cm,yshift=0.5cm,fill=orange!30,draw=black,text centered,minimum height=1.8cm,minimum width=0.8cm]{$U$};
    \node[xshift=-0.5cm,yshift=7cm,fill=white]{\scriptsize$q_0^r$};
    \node[xshift=-0.5cm,yshift=6cm,fill=white]{\scriptsize$q_1^g$};
    \node[xshift=-0.5cm,yshift=5cm,fill=white]{\scriptsize$q_2^b$};
    \node[xshift=-0.5cm,yshift=4cm,fill=white]{\scriptsize$q_3^o$};
    \node[xshift=-0.5cm,yshift=3cm,fill=white]{\scriptsize$q_4^r$};
    \node[xshift=-0.5cm,yshift=2cm,fill=white]{\scriptsize$q_5^g$};
    \node[xshift=-0.5cm,yshift=1cm,fill=white]{\scriptsize$q_6^b$};
    \node[xshift=-0.5cm,yshift=0cm,fill=white]{\scriptsize$q_7^o$};
    \node[xshift=4.5cm,yshift=7cm,fill=white]{\scriptsize$q_0^r$};
    \node[xshift=4.5cm,yshift=6cm,fill=white]{\scriptsize$q_4^r$};
    \node[xshift=4.5cm,yshift=5cm,fill=white]{\scriptsize$q_1^g$};
    \node[xshift=4.5cm,yshift=4cm,fill=white]{\scriptsize$q_5^g$};
    \node[xshift=4.5cm,yshift=3cm,fill=white]{\scriptsize$q_2^b$};
    \node[xshift=4.5cm,yshift=2cm,fill=white]{\scriptsize$q_6^b$};
    \node[xshift=4.5cm,yshift=1cm,fill=white]{\scriptsize$q_3^o$};
    \node[xshift=4.5cm,yshift=0cm,fill=white]{\scriptsize$q_7^o$};
    \end{tikzpicture}
    \end{minipage}
    \begin{minipage}{0.4\linewidth}
    \centering
    \scalebox{0.7}{
    \begin{tikzpicture}
    \draw[wire,red   ](-0.5,11)--(0.1,11)--(0.9,10)--(1.1,10)--(1.9, 9)--(2.1, 9)--(2.9, 9)--(3.1, 9)--(3.9, 8)--(4.1, 8)--(4.9, 8)--(5.5, 8);
    \draw[wire,red   ](-0.5, 5)--(0.1, 5)--(0.9, 6)--(1.1, 6)--(1.9, 7)--(2.1, 7)--(2.9, 8)--(3.1, 8)--(3.9, 9)--(4.1, 9)--(4.9, 9)--(5.5, 9);
    \draw[wire,green ](-0.5,10)--(0.1,10)--(0.9,11)--(1.1,11)--(1.9,11)--(2.1,11)--(2.9,10)--(3.1,10)--(3.9,10)--(4.1,10)--(4.9,10)--(5.5,10);
    \draw[wire,green ](-0.5, 8)--(0.1, 8)--(0.9, 9)--(1.1, 9)--(1.9,10)--(2.1,10)--(2.9,11)--(3.1,11)--(3.9,11)--(4.1,11)--(4.9,11)--(5.5,11);
    \draw[wire,blue  ](-0.5, 9)--(0.1, 9)--(0.9, 8)--(1.1, 8)--(1.9, 8)--(2.1, 8)--(2.9, 7)--(3.1, 7)--(3.9, 6)--(4.1, 6)--(4.9, 5)--(5.5, 5);
    \draw[wire,blue  ](-0.5, 1)--(0.1, 1)--(0.9, 2)--(1.1, 2)--(1.9, 3)--(2.1, 3)--(2.9, 4)--(3.1, 4)--(3.9, 5)--(4.1, 5)--(4.9, 6)--(5.5, 6);
    \draw[wire,violet](-0.5, 7)--(0.1, 7)--(0.9, 7)--(1.1, 7)--(1.9, 6)--(2.1, 6)--(2.9, 5)--(3.1, 5)--(3.9, 4)--(4.1, 4)--(4.9, 3)--(5.5, 3);
    \draw[wire,violet](-0.5, 0)--(0.1, 0)--(0.9, 0)--(1.1, 0)--(1.9, 1)--(2.1, 1)--(2.9, 2)--(3.1, 2)--(3.9, 3)--(4.1, 3)--(4.9, 4)--(5.5, 4);
    \draw[wire,orange](-0.5, 6)--(0.1, 6)--(0.9, 5)--(1.1, 5)--(1.9, 4)--(2.1, 4)--(2.9, 3)--(3.1, 3)--(3.9, 2)--(4.1, 2)--(4.9, 1)--(5.5, 1);
    \draw[wire,orange](-0.5, 2)--(0.1, 2)--(0.9, 1)--(1.1, 1)--(1.9, 0)--(2.1, 0)--(2.9, 0)--(3.1, 0)--(3.9, 1)--(4.1, 1)--(4.9, 2)--(5.5, 2);
    \draw[wire,cyan  ](-0.5, 4)--(0.1, 4)--(0.9, 3)--(1.1, 3)--(1.9, 2)--(2.1, 2)--(2.9, 1)--(3.1, 1)--(3.9, 0)--(4.1, 0)--(4.9, 0)--(5.5, 0);
    \draw[wire,cyan  ](-0.5, 3)--(0.1, 3)--(0.9, 4)--(1.1, 4)--(1.9, 5)--(2.1, 5)--(2.9, 6)--(3.1, 6)--(3.9, 7)--(4.1, 7)--(4.9, 7)--(5.5, 7);
    \node()[circle,xshift=0.5cm,yshift=1.5cm,fill=white,draw=black,text centered]{$f$};
    \node()[circle,xshift=0.5cm,yshift=5.5cm,fill=white,draw=black,text centered]{$f$};
    \node()[circle,xshift=0.5cm,yshift=8.5cm,fill=white,draw=black,text centered]{$f$};
    \node()[circle,xshift=0.5cm,yshift=10.5cm,fill=white,draw=black,text centered]{$f$};
    \node()[circle,xshift=1.5cm,yshift=0.5cm,fill=white,draw=black,text centered]{$f$};
    \node()[circle,xshift=1.5cm,yshift=2.5cm,fill=white,draw=black,text centered]{$f$};
    \node()[circle,xshift=1.5cm,yshift=4.5cm,fill=white,draw=black,text centered]{$f$};
    \node()[circle,xshift=1.5cm,yshift=6.5cm,fill=white,draw=black,text centered]{$f$};
    \node()[circle,xshift=1.5cm,yshift=9.5cm,fill=white,draw=black,text centered]{$f$};
    \node()[circle,xshift=2.5cm,yshift=1.5cm,fill=white,draw=black,text centered]{$f$};
    \node()[circle,xshift=2.5cm,yshift=3.5cm,fill=white,draw=black,text centered]{$f$};
    \node()[circle,xshift=2.5cm,yshift=5.5cm,fill=white,draw=black,text centered]{$f$};
    \node()[circle,xshift=2.5cm,yshift=7.5cm,fill=white,draw=black,text centered]{$f$};
    \node()[circle,xshift=3.5cm,yshift=0.5cm,fill=white,draw=black,text centered]{$f$};
    \node()[circle,xshift=3.5cm,yshift=2.5cm,fill=white,draw=black,text centered]{$f$};
    \node()[circle,xshift=3.5cm,yshift=4.5cm,fill=white,draw=black,text centered]{$f$};
    \node()[circle,xshift=3.5cm,yshift=6.5cm,fill=white,draw=black,text centered]{$f$};
    \node()[rectangle,xshift=3.5cm,yshift=8.5cm,fill=red!30,draw=black,text centered,minimum height=1.8cm,minimum width=0.8cm]{$U$};
    \node()[rectangle,xshift=2.5cm,yshift=10.5cm,fill=green!30,draw=black,text centered,minimum height=1.8cm,minimum width=0.8cm]{$U$};
    \node()[rectangle,xshift=4.5cm,yshift=5.5cm,fill=blue!30,draw=black,text centered,minimum height=1.8cm,minimum width=0.8cm]{$U$};
    \node()[rectangle,xshift=4.5cm,yshift=3.5cm,fill=violet!30,draw=black,text centered,minimum height=1.8cm,minimum width=0.8cm]{$U$};
    \node()[rectangle,xshift=4.5cm,yshift=1.5cm,fill=orange!30,draw=black,text centered,minimum height=1.8cm,minimum width=0.8cm]{$U$};
    \node()[rectangle,xshift=0.5cm,yshift=3.5cm,fill=cyan!30,draw=black,text centered,minimum height=1.8cm,minimum width=0.8cm]{$U$};
    \node[xshift=-0.5cm,yshift=11cm,fill=white]{\scriptsize$q_0^r$};
    \node[xshift=-0.5cm,yshift=10cm,fill=white]{\scriptsize$q_1^g$};
    \node[xshift=-0.5cm,yshift=9cm,fill=white]{\scriptsize$q_2^b$};
    \node[xshift=-0.5cm,yshift=8cm,fill=white]{\scriptsize$q_3^g$};
    \node[xshift=-0.5cm,yshift=7cm,fill=white]{\scriptsize$q_4^p$};
    \node[xshift=-0.5cm,yshift=6cm,fill=white]{\scriptsize$q_5^o$};
    \node[xshift=-0.5cm,yshift=5cm,fill=white]{\scriptsize$q_6^r$};
    \node[xshift=-0.5cm,yshift=4cm,fill=white]{\scriptsize$q_7^c$};
    \node[xshift=-0.5cm,yshift=3cm,fill=white]{\scriptsize$q_8^c$};
    \node[xshift=-0.5cm,yshift=2cm,fill=white]{\scriptsize$q_9^o$};
    \node[xshift=-0.5cm,yshift=1cm,fill=white]{\scriptsize$q_{10}^b$};
    \node[xshift=-0.5cm,yshift=0cm,fill=white]{\scriptsize$q_{11}^p$};
    \node[xshift=5.5cm,yshift=11cm,fill=white]{\scriptsize$q_1^g$};
    \node[xshift=5.5cm,yshift=10cm,fill=white]{\scriptsize$q_3^g$};
    \node[xshift=5.5cm,yshift=9cm,fill=white]{\scriptsize$q_0^r$};
    \node[xshift=5.5cm,yshift=8cm,fill=white]{\scriptsize$q_6^r$};
    \node[xshift=5.5cm,yshift=7cm,fill=white]{\scriptsize$q_7^c$};
    \node[xshift=5.5cm,yshift=6cm,fill=white]{\scriptsize$q_2^b$};
    \node[xshift=5.5cm,yshift=5cm,fill=white]{\scriptsize$q_{10}^b$};
    \node[xshift=5.5cm,yshift=4cm,fill=white]{\scriptsize$q_4^p$};
    \node[xshift=5.5cm,yshift=3cm,fill=white]{\scriptsize$q_{11}^p$};
    \node[xshift=5.5cm,yshift=2cm,fill=white]{\scriptsize$q_5^o$};
    \node[xshift=5.5cm,yshift=1cm,fill=white]{\scriptsize$q_9^o$};
    \node[xshift=5.5cm,yshift=0cm,fill=white]{\scriptsize$q_8^c$};
    \end{tikzpicture}
    }
    \end{minipage}
    \end{center}
    \caption{Example circuits for logical-qubit routing from the integer linear program for 8 (left) and 12 (right) qubits.
    The problem is to interact each qubit with its partner (matching color/superscript) in as few timesteps as possible, with a secondary objective to minimise the number of swaps.
    The $f$ gates are fermionic swaps while the $U$ gates are the interactions given in Figure~\ref{fig:G2} for converting the joint measurement into single qubit measurements.}
    \label{fig:routing}
\end{figure*}
}
\newcommand{\includefigcalibration}[1]{
\begin{figure*}[#1]
    \centering
    \includegraphics[width=0.7\linewidth]{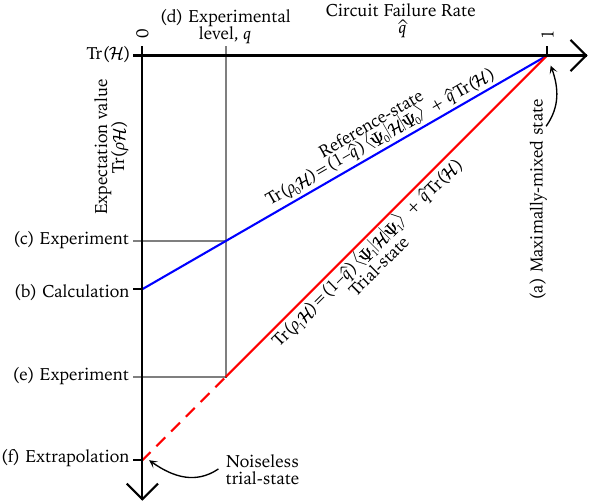}
    \caption{The reference-state calibration method can be visualised on a plot of expectation value vs. circuit failure rate.
    The maximally noisy (a) and noiseless (b) energies of the reference-state are classically tractable, so using quantum computation of the reference-energy (c) allows estimation of the effective noise level in the device (d).
    By evaluating the energy of the trial-state on a quantum computer (e) the noiseless energy of the trial-state (f) can be estimated by extrapolation.
    The linear relationship is due to the white-noise model with more sophisticated noise models requiring more reference-state calculations to fit~\cite{Czarnik_CDR_2021}.
    The reference-state error mitigation scheme differs from the usual zero-noise extrapolation method~\cite{Temme_ZNE_2017,Kandala_ZNE_2019} since the additional information is obtained through reference-states instead of artificially increasing the noise.
    }
    \label{fig:calibration}
\end{figure*}
}

\begin{document}
% \title{Electronic structure of water using quantum computed moments}
% \title{Precision ground-state energy estimation of the water molecule on a physical quantum computer}
\title{Precision ground-state energy calculation for the water molecule on a superconducting quantum processor}
\authors{
Michael A. Jones,
Harish J. Vallury,
Lloyd C. L. Hollenberg
}{
School of Physics, University of Melbourne, Parkville 3010, AUSTRALIA
}
\abstract{
The accurate computation of properties of large molecular systems is classically infeasible and is one of the applications in which it is hoped that quantum computers will demonstrate an advantage over classical devices. However, due to the limitations of present-day quantum hardware, 
variational-hybrid algorithms introduced to tackle these problems struggle to meet the accuracy and precision requirements of chemical applications.
Here, we apply the Quantum Computed Moments (QCM) approach combined with a variety of noise-mitigation techniques to an 8 qubit/spin-orbital representation of the water molecule (H$_\mathbf{2}$O).
A noise-stable improvement on the variational result for a 4-excitation trial-state (circuit depth 25, 22 CNOTs) was obtained, with the ground-state energy computed to be within $\mathbf{1.4\pm1.2}$ mHa of exact diagonalisation in the 14 spin-orbital basis. Thus, the QCM approach, despite an increased number of measurements and noisy quantum hardware (CNOT error rates c.1\% corresponding to expected error rates on the trial-state circuit of order 20\%), is able to determine the ground-state energy of a non-trivial molecular system at the required accuracy (c.0.1\%). To the best of our knowledge, these results are the largest calculations performed on a physical quantum computer to date in terms of encoding individual spin-orbitals producing chemically relevant accuracy, and a promising indicator of how such hybrid approaches might scale to problems of interest in the low-error/fault-tolerant regimes as quantum computers develop.
}

\begin{multicols}{2}
\section{Introduction}\label{sec:intro}
While quantum computing promises to revolutionise computational quantum chemistry in the long-term (i.e. using millions of error-corrected qubits,~\cite{Kitaev_QPE_1995,ReiherWiebe_FeMoco_2017}), there is also considerable interest in whether near-term quantum devices can provide an advantage for chemistry applications.
In this Noisy Intermediate-Scale Quantum~\cite{Preskill_NISQ_2018} regime, the mainstay approach to computing quantities such as the ground-state energy, $E_0$, of molecular systems defined by a Hamiltonian, $\mathcal{H}$, are hybrid quantum-classical algorithms
\cite{PeruzzoMcClean_VQE_2014,Cerezo_VQAs_2021} based on the variational principle:
\begin{align}
    \langle\mathcal{H}\rangle=\langle\Psi_\mathrm{trial}|\mathcal{H}|\Psi_\mathrm{trial}\rangle\ge E_0, \label{eqn:variational}
\end{align}
The expectation is that, as the problem size scales, the preparation and measurement of the (parameterised) trial-state, $|\Psi_\mathrm{trial}\rangle$, on the quantum device will be less expensive than for the classical counterpart.
Variational algorithms have been successfully applied to relatively small ($\le6$ qubit) systems~\cite{
PeruzzoMcClean_VQE_2014,
OMalleyBabbush_VQE_2016,
KandalaMezzacapo_IBM_VQE_2017,
Shen_VQE_2017,
Colless_QSE_2018,
Ganzhorn_VQE_2019,
Kandala_ZNE_2019,
McCaskeyJakowskiPooser_AlkaliHydrides_2019,
Sagastizabal_SV_2019,
Smart_NRepresentability_2019,
Nam_IonQWater_2020,
Smart_TwoElectron_2020,
Gao_Lithium_2021,
Kawashima_H10_2021,
Rice_Batteries_2021,
Tilly_QRDM_2021,
Eddins_EntanglementForging_2022,
Jones_Hydrogen_2022,
Shee_ConfigurationEncoding_2022,
Suchsland_VariationalMoments_2021,
Yamamoto_Periodic_2022,
GuoSunQianGong_USTC_2023,
Khan_ChemicallyAwareUCC_2023,
LolurSkogh_ReferenceStates_2023,
Motta_EntanglementForging_2023,
RossmannekPavosevic_Embedding_2023,
Shirai_SpinFree_2023,
Yoffe_ConfigurationEncoding_2023}
and/or restricted trial-states (e.g. Hartree-Fock~\cite{Google_HF_2020,Jones_Hydrogen_2022} or pair-correlated~\cite{OBrien_UpCC_2022,Zhao_UpCC_2023} states), with recent work \cite{Goings_Benzene_2023} extending to to higher qubit counts.
The key issue preventing scale-up of the approach is the effect of device noise on the trial-state preparation, which in the context of required accuracy rapidly overwhelms the computation, pushing the variational upper bound away from the true ground-state energy.

A promising improvement on the variational approach proposed recently is to use a noise robust ground-state energy estimate from Lanczos expansion theory~\cite{Hollenberg_PlaquetteExpansion_1993,
HollenbergWitte_Nonperturbative_1994,
HollenbergWitte_AnalyticSolution_1996,
HollenbergBardosWitte_ClusterExpansion_1996} based on higher order Hamiltonian moments, $\langle\mathcal{H}^p\rangle$.
The Quantum Computed Moments (QCM,~\cite{Vallury_QCM_2020}) approach has been seen to improve on the variational energy estimate in the presence of noise~\cite{Vallury_NoiseRobust_2023} and has been applied to hydrogen atom chains for restricted trial-states up to 6 qubits~\cite{Jones_Hydrogen_2022}.
This work seeks to extend the application of the QCM approach in both number of qubits and trial-state complexity by computation of the ground-state energy of
the water molecule (H$_2$O) using a Unitary Coupled Cluster Doubles (UCCD,~\cite{Taube_UCC_2006}) based ans\"atz defined over 8 qubits/spin-orbitals on a superconducting quantum processor through the workflow presented in Figure~\ref{fig:flow}.
The trial-circuits used encode each spin-orbital into a separate qubit and implement 4 and 6 double excitations requiring 22 and 72 CNOT gates respectively (Figures~\ref{fig:circuits}b and c).
Such an encoding is preferable to assigning one spatial-orbital (2 spin-orbitals) per qubit, as it allows for more flexibility in trial-state design.
On the other hand, encoding spatial-orbitals allows for much more compact trial-states, even for the same number of qubits.

Here, we show that through the use of a carefully designed trial-circuit, noise mitigation techniques and the QCM approach, an error from exact diagonalisation in the minimal STO-3G basis (14 spin-orbitals) of 1-3 mHa is achieved for the 4-excitation trial-state circuit.
To the best of our knowledge this is one of the largest quantum computed electronic structure calculations (in terms of the number of entangled qubits) that encodes each spin-orbital individually, and achieves chemically relevant accuracy at the mHa level.
Additionally, the computations are performed in a gate-based framework and do not employ techniques such as pulse optimisation or long-term device monitoring.

The remainder of this paper is arranged as follows: 
Section~\ref{sec:background} provides a brief background on the relevant quantum chemistry, the Variational Quantum Eigensolver (VQE,~\cite{PeruzzoMcClean_VQE_2014}) and the QCM method.
Section~\ref{sec:results} presents the results obtained in this work and Section~\ref{sec:discussion} briefly discusses the significance of these results.
Section~\ref{sec:methods} summarises the methods employed in this work with additional detail provided in the supplementary information.

\section{Background}\label{sec:background}
\includefigflow{p}
\includefigcircuit{t}
\subsection{Quantum chemistry}
Given a fixed nuclear geometry, the electronic structure Hamiltonian of a molecular system can be written in second-quantisation as:
\begin{align}
    \mathcal{H}&=\sum_{jk}h_{jk}a^\dagger_ja_k+\sum_{jklm}g_{jklm}a^\dagger_ja^\dagger_ka_la_m, \label{eqn:hamiltonian}
\end{align}
where $a^\dagger_j$ ($a_j$) are the creation (annihilation) operators for an electron in spin-orbital $j$, and $h_{jk}$ ($g_{jklm}$) are the classically computed 1- (2-) body electronic integrals defined by the chosen basis of spin-orbitals (Figure~\ref{fig:flow}a,b).
Implicit in this definition is the discretisation of space into a basis set of 3-dimensional electronic wavefunctions, known as orbitals.
Each orbital is composed of two spin-orbitals (corresponding to the spin degree of freedom of a single electron) and this discretisation of space introduces a level of approximation to the problem, with the error vanishing as the number of spin-orbitals approaches infinity.
Given the Hamiltonian, the problem is then to estimate the ground-state energy to high accuracy.
The ultimate goal of quantum chemistry methods is to achieve `chemical accuracy', which is defined to be a result within 1.6 mHa of the infinite basis limit \emph{or} within the same margin of accurate experimental methods.
A slightly less demanding target is to achieve `FCI accuracy', where a result is within 1.6mHa of Full Configuration Interaction (FCI, equivalent to matrix diagonalisation) for a given basis size.
FCI accuracy has been achieved on noisy quantum devices for various small systems using up to 5-6 
qubits~\cite{
Colless_QSE_2018,
KandalaMezzacapo_IBM_VQE_2017,
McCaskeyJakowskiPooser_AlkaliHydrides_2019,
Kawashima_H10_2021,
Tilly_QRDM_2021,
Eddins_EntanglementForging_2022,
Jones_Hydrogen_2022,
GuoSunQianGong_USTC_2023,
Motta_EntanglementForging_2023,
Shirai_SpinFree_2023,
Yoffe_ConfigurationEncoding_2023}.
Finally, `trial-state accuracy', can be defined as when the noisy quantum results reproduce noiseless simulation of the trial-state to within 1.6 mHa~\cite{Google_HF_2020,Nam_IonQWater_2020}.
The results presented here achieve FCI accuracy, though the FCI computation is performed over 14 spin-orbitals while the quantum computation is performed over 8 spin-orbitals. Of the remaining 6 orbitals, 2 core orbitals are neglected completely and the 4 virtual orbitals are incorporated via the QCM approach.
\subsection{The variational quantum eigensolver method}
Expressing the electronic structure problem as a second-quantised Hamiltonian allows for a direct mapping between spin-orbital occupation states and qubit computational basis states, i.e. the Jordan-Wigner encoding~\cite{JordanWigner_JWE_1928}.
In this way each spin-orbital corresponds to one logical qubit (these are referred to as logical qubits because the information may be swapped between physical qubits) and the problem is to construct a state, $|\Psi_\mathrm{trial}\rangle=|\Psi(\vec\theta)\rangle$, on the quantum computer and identify the values of $\vec\theta$ for which the energy is minimised (Figure \ref{fig:flow}c).
The possibility of quantum advantage comes from the fact that, unlike classical computers which require memory scaling exponentially with the number of spin-orbitals, a quantum computer requires only a number of qubits linear in the number of spin-orbitals to represent an arbitrary trial-state.
While completely general trial-states will require exponentially deep circuits, it is possible to construct reasonable, physically motivated trial-states in polynomial depth such as via the (Trotterised) UCC operator as discussed in Section~\ref{sec:methods} and~\ref{app:trialstate}.
Once the parameterised quantum circuit is defined, the energy can be measured efficiently and the parameters optimised via a classical minimisation loop~\cite{PeruzzoMcClean_VQE_2014}.
From Equation~\ref{eqn:variational}, this minimum is an upper-bound to the true ground-state energy and, assuming no hardware noise, that the trial-circuit is expressive enough, and the optimisation has been performed well, it will be a tight upper-bound.

\subsection{The quantum computed moments method}
The theory underpinning the energy correction used in the QCM method~\cite{Vallury_QCM_2020} is related to quantum subspace expansion, quantum Krylov methods and other quantum moments-based methods~\cite{
McClean_QSE_2017,
Colless_QSE_2018,
Takeshita_QSE_2020,
Motta_QLanczos_2020,
Yeter-Aydeniz_QITEQLanczos_2020,
Huggins_NOVQE_2020,
Parrish_QFD_2019,
Cohn_QFD_2021,
Stair_MRSQK_2020,
Cortes_QKrylov_2022,
Seki_QuantumPowerMethod_2021,
SekiYunoki_FermiHubbard_2022,
Suchsland_VariationalMoments_2021,
Kowalski_CMX_2020,
Peng_VariationalPDS_2021,
Claudino_CMX_2021,
Claudino_SingletFission_2023,
Kirby_Lanczos_2023,
Klymko_TimeEvolution_2022}, that effectively generate a reduced Hamiltonian in a subspace that (ideally) includes the ground-state.
The implementation of the QCM method used here employs an analytic form for the corrected ground-state energy, obtaned from Lanczos cluster expansion theory \cite{Hollenberg_PlaquetteExpansion_1993}, to fourth order in the Hamiltonian moments, $\langle\mathcal{H}^p\rangle$~
\cite{HollenbergWitte_Nonperturbative_1994}:
\begin{align}
    E_L&=c_1-\frac{c_2^2}{c_3^2-c_2c_4}\left(\sqrt{3c_3^2-2c_2c_4}-c_3\right),\label{eqn:EL}
\end{align}
where the cumulants are given by
\begin{align}
    c_p&=\langle\mathcal{H}^p\rangle-\sum_{j=0}^{p-2}\left(\mat{p-1\\j}\right)c_{j+1}\langle \mathcal{H}^{p-1-j}\rangle.\label{eqn:cumulants}
\end{align}
Here, expectation values are taken with respect to the optimised trial-state.
In addition to avoiding the explicit construction and diagonalisation of the reduced Hamiltonian, Equation~\ref{eqn:EL} has been seen to be surprisingly robust against noise~\cite{Vallury_NoiseRobust_2023} in the computation of the moments compared to evaluation of $\langle\mathcal{H}\rangle$ alone.
Another possible application is the extension of the QCM method to the computation of Green's functions as demonstrated recently \cite{GreenDiniz_GreensFunctions_2023}.
In the context of chemical computations, the QCM method has been applied to electronic structure problems using single Slater-determinant states~\cite{Jones_Hydrogen_2022}.
Here we employ multi-determinant trial-states which require much deeper trial-circuits and a range of error mitigation techniques.

\section{Results}\label{sec:results}
\includefigres{t}
Calculations were performed following the overall procedure outlined in Figure~\ref{fig:flow}.
The Trotterised UCCD trial-circuit was optimised using a classical statevector simulator; the problem of optimising the trial-state in the presence of noise is left to future work.
The final computations were performed on the \emph{ibmq\_kolkata} device.
Additional method details can be found in Section~\ref{sec:methods} and the supplementary information.

Two different trial-state depths were used, the state preparation circuit in Figure~\ref{fig:results}a has a depth of 25 (22 CNOT gates) and implements 4 excitations from the truncated Unitary Coupled Cluster ans\"atz (see~\ref{app:trialstate}) while the circuit in Figure~\ref{fig:results}b has depth 74 (72 CNOT gates) and implements 6 excitations.
Note that these depths and gate counts include only the state-preparation component of the circuit and that different measurement bases increase these values by different amounts.
The results for both energy estimates $\langle\mathcal{H}\rangle$ and $E_L$ and for both trial-states are presented in Figure~\ref{fig:results}.
An important observation regarding the results presented is that the moments of the Hartree-Fock state are classically tractable (and are required for the reference-state calibration technique used for noise mitigation, see~\ref{app:calibration}).
Therefore, if the quantum computer is going to provide any utility in the computation, it must give expectation values that are (at least) competitive with this reference-state energy (black dashed lines in Figure~\ref{fig:results}).
It is also important to note that the energy computed from Equation~\ref{eqn:EL} is non-variational, so the FCI energy is not a lower bound which can lead to `negative' errors.
\subsection{4-excitation trial-state}
For the 4-excitation trial-state, both energy estimates ($\langle\mathcal{H}\rangle$ and $E_L$) have lower error than the reference-state.
$\langle\mathcal{H}\rangle$ in particular, is separated from the corresponding reference-state estimate by more than 2 standard deviations, indicating that the benefits of implementing the quantum circuit outweigh the cost (in accuracy) imposed by device noise.
For $E_L$, the separation between noise-mitigated and reference-state energies (in absolute error from FCI) is less than half a standard deviation.
The reduced resolution can be largely attributed to the high accuracy of the moments method applied to the reference-state, which has an error only 0.5mHa larger than the statevector simulation.
Given that the target accuracy is 1.6mHa, the obtained standard deviation of 1.2mHa is acceptable and it is unnecessary to attempt to reduce this further.
Instead, it may be of interest to repeat the experiment for a system that is not well modelled by Hartree-Fock methods to determine if the resolution between reference and noise-mitigated energies can be increased.
\subsection{6-excitation trial-state}
Due to the significant depth increase to the trial-circuit, the 6-excitation trial-state energies have greater error than the reference-state energies.
Of particular interest is that the $\langle\mathcal{H}\rangle$ energies differ by more than 2 standard deviations.
When considering the reference-state calibration, this can only be possible when (in the presence of noise) the reference-state is found to have lower energy than the trial-state, suggesting the possibility of errors not captured by the white-noise assumption.
A possible method to mitigate this would be to use randomised compiling~\cite{Hashim_RandomisedCompiling_2021} to convert unitary and/or biased measurement noise into unbiased stochastic noise.
Alternatively, the unexpected behaviour of the energy may be due to a failure to satisfy the $N$-representability conditions that determine whether the computed reduced density matrices are physical (see~\ref{app:rescaling}).

\section{Discussion}\label{sec:discussion}
While larger electronic structure computations have been performed, these are made possible by encoding multiple spin-orbitals into a single qubit~\cite{Zhao_UpCC_2023,OBrien_UpCC_2022,Google_HF_2020,Kawashima_H10_2021} and/or through exploiting the assumption of weak entanglement~\cite{Eddins_EntanglementForging_2022,Motta_EntanglementForging_2023}.
This can allow simplification of paired double-excitations to circuit elements involving only two CNOT gates, however, it prohibits the implementation of single- and unpaired double- excitations, limiting the effectiveness of the trial-state.
Here, knowledge of the initial state of the circuit is used to simplify the implementation of the first 4 excitations, however, the circuit is not restricted to paired coupled cluster states.
After these 4 simplified excitations, the general form of the double excitation gate is required, which results in the comparatively deep 6-excitation circuit.

The accuracy obtained here for the water molecule compares well with the accuracy achieved for the benzene molecule using VQE for 8 qubits and advanced circuit optimisation on a trapped-ion quantum computer \cite{Goings_Benzene_2023}, though since the computations are performed for different molecules, a direct comparison is difficult. Additionally, the trapped-ion energy errors are reported relative to 8 spin-orbital complete active space calculations, while we compare to the more accurate 14 spin-orbital FCI. Notably, the trapped-ion experiment required fewer state preparations at the expense of deeper circuits. This trade-off is favourable for trapped-ion qubits which have longer coherence times but a lower repetition rate than superconducting qubits.

We demonstrate that up to circuit depths~$>25$ and widths of 8 qubits, there is sufficient information contained in the output probability distributions from IBM Quantum hardware to outperform the classical reference-state.
It is important to note that the energy scale on which this comparison takes place is much smaller than the scale set by the maximally mixed state, which has error on the order of 1-2 Ha (500-800 mHa) for $\langle\mathcal{H}\rangle$ ($E_L$). This is 1-2 orders of magnitude larger than the states of interest.
This result demonstrates that it is possible for moments-based methods to extend the reach of variational algorithms for quantum chemistry on present-day and near-term quantum hardware, pointing the way towards a potential for quantum utility.

\section{Methods}\label{sec:methods}
\subsubsubsection{Electronic Hamiltonian}
The electronic Hamiltonian is calculated using the \emph{PySCF}~\cite{Sun_PySCF_2017} python package in the minimal STO-3G basis comprising 14 spin-orbitals.
The core orbitals (predominantly the 1s orbitals of the O atom) are frozen and the Hamiltonian moments are computed via Wick’s theorem in the 12 spin-orbital basis.
The Hamiltonian and its moments are then reduced further, to 8 spin-orbitals by freezing the 4 orbitals that participate least in the bonding.
By computing moments in the 12 spin-orbital basis, the moments-corrected energies are able to extend the computation into the otherwise unused virtual orbitals.
The FCI energies are also computed by \emph{PySCF} in the 14 spin-orbital basis.

\subsubsubsection{Trial-circuit}
The trial-circuit is based on the Trotterisation of the chemically-inspired Unitary Coupled Cluster Doubles (UCCD) ans\"atz with the double excitations ordered based on their known contribution to the ground-state.
The circuit is then compiled to enforce linear connectivity between qubits and to respect the gateset native to the quantum device.
See~\ref{app:trialstate} for details.

\subsubsubsection{Measurement bases}
Using approximately 200 fermionic measurement bases allows computation of the 940 non-trivial elements of the 4-body reduced density matrix.
Each correlated excitation is decomposed to a product of uncorrelated excitations and these are grouped using a greedy algorithm.
The measurement circuits are optimised by solving an integer linear program using \emph{Gurobi}~\cite{Gurobi_2023}.
See~\ref{app:RDM}.

\subsubsubsection{Energy minimisation}
Energy minimisation is performed using a customised SPSA optimiser on noiseless, classical simulation.
Minimisation was performed 5 times using different seeds for the stochastic optimiser and the optimised parameters from the best run were used for the hardware experiment, though the difference between runs was minimal.
The optimisation generally converged after 50-60 iterations.

\subsubsubsection{IBM quantum backends}
The quantum computations in this work were performed on \emph{ibmq\_kolkata}, one of IBM's superconducting quantum devices~\cite{Qiskit_2019}, using 8 of the 27 available qubits.
Each measurement basis was measured $10^5$ times; for 200 measurement bases at two different values of $\vec{\theta}$ (trial- and reference-states) and including the readout-noise calibration, this leads to approximately 42 million shots and a total quantum computation time of 1.5 hours (per trial-state).
Uncertainty estimation is performed by a statistical bootstrapping technique~\cite{EfronTibshirani_Bootstrapping_1994}.
See~\ref{app:bootstrapping}.

\subsubsubsection{Noise mitigation}
Scalable readout-noise mitigation, symmetry verification by post selection on the electron number and total spin, reduced density matrix rescaling and reference-state calibration are implemented to mitigate noise accumulated during the quantum computation.
See~\ref{app:mitigation}.
In addition the QCM method is able to partially account for the limited expressivity of the trial-state and is robust to the remaining noise.
See references~\cite{Vallury_QCM_2020,Jones_Hydrogen_2022,Vallury_NoiseRobust_2023}.

\section{Competing Interests}
The QCM method is the subject of an international patent application no PCT/AU2021/050674.
\section{Data Availability}
The data that support the findings of this study are available from the corresponding author upon reasonable request.
% note to self: data is stored locally, code is on Spartan under home directory.
\section{Author Contributions}
LCLH conceived the project. MAJ set up the computational framework and performed the calculations and data analysis, with input from all authors.
\section{Acknowledgements}
The research was supported by the University of Melbourne through the establishment of the IBM Q Network Hub at the University.
MAJ and HJV are supported by the Australian Commonwealth Government through Research Training Program Scholarships. 
The authors would like to thank G. Mooney, H. Quiney and A. Martin for useful discussions.
This work was supported by resources provided by the Pawsey Supercomputing Research Centre with funding from the Australian Government and the Government of Western Australia, %Pawsey
and by The University of Melbourne’s Research Computing Services and the Petascale Campus Initiative. %Spartan

\end{multicols}
\references{bibliography}

\supinf
\begin{multicols}{2}
\section{Trial-state design}\label{app:trialstate}
\includefigfswap[0.7]{p}
\includefigGfour[0.7]{p}
One of the most challenging aspects of implementing variational hybrid algorithms is choosing a trial-circuit that is expressive enough to contain (a good approximation to) the true ground-state and at the same time is shallow enough for the quantum device to remain coherent throughout the state preparation.
In addition, care should be taken to avoid over-parameterisation of the ans\"atz which can lead to so-called ``barren plateaus''\cite{McClean_Plateaus_2018} -- regions of exponentially vanishing gradient that cause classical optimisation to become infeasible.
A common, physically motivated, starting point for creating trial-circuits for quantum chemistry is the Unitary Coupled Cluster (UCC,~\cite{Taube_UCC_2006}) ans\"atz:
\begin{align}
    |\Psi_\mathrm{UCC}\rangle&=e^{i\tau}|\Psi_\mathrm{init}\rangle,\\
    \tau&=i(T-T^\dagger)\nonumber
\end{align}
where $T$ is a weighted sum over all possible electronic excitations, including both uncorrelated (single) and correlated (double, triple, etc.) excitations.
The initial state, $|\Psi_\mathrm{init}\rangle$, is an easily prepared state in the correct symmetry sector of Hilbert space and is usually taken to be the Hartree-Fock state, $|\Psi_\mathrm{HF}\rangle$.
While the UCC trial-state is expected to give a good approximation of the true ground-state as the number of excitations is increased (in the limit that all excitations are included the trial-state can exactly express the ground state), practical implementation requires both limitation to some maximum excitation order and Trotterisation.
Here, we use the double-excitation cluster operator, $T_2$, and Trotterise to first order.
\begin{align}
    |\Psi_\mathrm{trial}\rangle&=\prod_{jklm}G_{jklm}(\theta_{jklm})\cdot|\Psi_\mathrm{HF}\rangle,\n
    G_{jklm}(\theta)&=\exp\left(\theta(a^\dagger_ja^\dagger_ka_la_m-a^\dagger_ma^\dagger_la_ka_j)\right)    
\end{align}
where $a^\dagger_ja^\dagger_ka_la_m$ implements the electronic double-excitation (the hermitian conjugate ensures unitarity of the $G$ operator) and $\theta_{jklm}$ are variational parameters (corresponding to the weights in the original cluster operator, $T$).
Given sufficient circuit depth, the single- and double- excitation ans\"atz has been seen to be an effective trial-state~\cite{Chen_FlexibleUCC_2022}.
In this case the exclusion of single-excitation operators is partially justified by the observation that leading single-excitations cannot introduce multi-reference character and therefore cannot improve on the Hartree-Fock state, while trailing single-excitations can be accounted for efficiently with classical processing~\cite{Thouless_HartreeFock_1960,Sokolov_OrbitalOptimisation_2020,Zhao_UpCC_2023}.
In the Trotterisation of the cluster operator, there is a degree of freedom as to the order that non-commuting component excitations should be performed in.
Here, the order is chosen based on the known contribution of each excitation term to the true ground-state~\cite{Nam_IonQWater_2020} and, to limit circuit depth and permit implementation on present-day devices, only the first 4-6 excitations are considered.
In practice this ordering would not be known \emph{a priori}, however, it should be possible to make use of physical/chemical intuition and/or an adaptive ans\"atz~\cite{Grimsley_Adapt_2019,Tang_QubitAdapt_2021} to determine a suitable ordering.
This choice of ordering further justifies the lack of single-excitations, as they are seen to have a smaller contribution than double-excitations to the true ground-state for the specific Hamiltonian considered here.

Since the trial-state preparation operator is now written as a product of double-excitations, a quantum circuit to implement the operator can be separated into a number of `blocks' each of which implements a double-excitation operation on a set of four fermionic modes.
By considering the fermionic modes as logical qubits on a linear array of physical qubits it can be shown that using the fermionic swap operator to permute the involved fermionic modes together converts the highly non-local excitation to a local interaction (shown in Figure~\ref{fig:fswap}).
Decomposition of the local fermionic interaction leads to 19 CNOT gates and 19 single-qubit ($H$, $S$, $S^\dagger$ and $R_y(\theta)$) gates with linear connectivity as shown in Figure~\ref{fig:G4}.

\section{Reduced density matrices}\label{app:RDM}
Generally, measurement within the VQE framework involves forming mutually commuting partitions of Hamiltonian fragments~\cite{KandalaMezzacapo_IBM_VQE_2017}.
An alternative is to obtain a classical description of the quantum state from which relevant operators can be estimated, such as via `classical shadows' ~\cite{Huang_ClassicalShadows_2020,Zhao_FermionicShadows_2021} or similar techniques.
Here, the 4-body Reduced Density Matrix (4-RDM) of the trial-state is measured:
\begin{align}
    R_{jklm}^{j'k'l'm'}&=\langle a^\dagger_ja^\dagger_ka^\dagger_la^\dagger_ma_{j'}a_{k'}a_{l'}a_{m'}\rangle. \label{eqn:rdm}
\end{align}
The RDM is an efficient classical description of the state from which expectation values of lower order excitations can be reconstructed:
\begin{align}
    R_{j_1\dots j_p}^{k_1\dots k_p}&=\frac{1}{N_e-p}\sum_{l}R_{j_1\dots j_pl}^{k_1\dots k_pl}.\label{eqn:reconstruction}
\end{align}
Therefore, since any second quantised electronic structure Hamiltonian can be written in the form of Equation~\ref{eqn:hamiltonian}, measurement of the 2-RDM allows computation of \emph{any} molecular energy (for a given trial-circuit).
Additionally, other properties of interest can be written as sums of low order excitation operators and hence evaluated from the RDM.
For example: the dipole moment is a sum over single-excitation operators
\begin{align}
    \mu&=\sum_{jk}t_{jk}\langle a^\dagger_ja_k\rangle=\sum_{jk}t_{jk}R_j^k.
\end{align}
The choice, in this case, to measure the 4-RDM is due to the 4th Hamiltonian moment containing up to 8-body excitations and therefore requiring estimation of the 8-RDM in general.
However, due to the fact that there are only 4 electrons present in the trial-state, excitation operators beyond 4th order will have vanishing expectation values.

Below we present a method for grouping and measuring correlated excitation operators for the estimation of reduced density matrices.

\subsection{Filtering RDM elements}\label{app:filtering}
In general the number of $p$-RDM elements for $N_s$ spin-orbitals is
\begin{align}
    N_r&=\frac{1}{2}\left(\frac{N_s!}{(N_s-p)!p!}\right)^2=\mathcal{O}({N_s}^{2p}),
\end{align}
where it can be shown that 
\begin{itemize}
\item repeated indices in either the superscript or subscript on the left-hand side of Equation~\ref{eqn:rdm} lead to the right-hand side vanishing,
\item permutations of indices only introduce factors of $-1$ due to fermionic anti-commutation relations,
\item interchanging upper and lower indices is equivalent to complex conjugation.
\end{itemize}
For the 4-RDM considered here, this is 2485 operators but can be reduced further by considering spin-symmetry~\cite{Tilly_QRDM_2021}.
Since the trial-state is constructed within a specific spin-subspace of the full Hilbert space, any excitation that does not conserve spin must have vanishing expectation value.
This restriction reduces the number of operators requiring measurement to 940, which (combined with the grouping method introduced below) is enough reduction to allow computation within a reasonable time.
\subsection{Measurement of uncorrelated excitations}
\includefigGtwo{t}
Before generalising to correlated excitations, we first seek to measure uncorrelated excitations, i.e. operators of the form $a^\dagger_ja_k$.
One method to do this involves applying the Jordan-Wigner transformation to convert the operator to a sum of Pauli-strings and measuring each string separately.
Unfortunately, this can lead to highly non-local strings which are less likely to qubit-wise commute and are more prone to bit-flip errors in the readout process.
Additionally, it is not possible to perform post selection when measuring Pauli-strings as they do not necessarily commute with the total number operator or the spin-projection operator.
Instead, consider the case $k=j\pm1$.
In this case, the excitation operator is a 2-local operator and its real and imaginary components are diagonalised (and hence measured) by the gate sequences given in Figure~\ref{fig:G2}.
For uncorrelated excitations on real-valued spin-orbitals, the imaginary component is guaranteed to have vanishing expectation but will be required when evaluating correlated excitations.
In the case that $k\notin\{j,j\pm1\}$, the fermionic swap operation (Figure~\ref{fig:fswap}) can be used to reorder the logical qubits so that they are adjacent.
Finally, if $j=k$ then the operator is the number operator and its eigenvalue is simply the bit-value when measured in the computational basis.

A disadvantage of this measurement scheme is the necessity of fermionic swap operations which extends the circuit depth by $\mathcal{O}(N_s)$.
On the other hand it avoids non-local measurement strings and allows symmetry verification by post-selection on the total electron number (since all circuits in Figure~\ref{fig:G2} commute with the total electron number operator).
Additionally, the only single excitations that do not commute with the total spin-projection operator are those for which $j$ and $k$ have different spins.
From the arguments made in~\ref{app:filtering}, these operators vanish in expectation and their measurement is therefore unnecessary, allowing post selection based on the total spin-projection.

\subsection{Measurement of correlated excitations}
Measuring correlated excitations directly is more difficult than measuring uncorrelated ones as the circuit depth required for diagonalisation increases with excitation order.
Instead, it is possible to decompose a correlated excitation into a sum of products of real and imaginary components of uncorrelated excitations, e.g. assuming unique subscripts:
\begin{align}
    a^\dagger_{j_1}a^\dagger_{j_2}a_{k_1}a_{k_2}&=
    -(a^\dagger_{j_1}a_{k_1})(a^\dagger_{j_2}a_{k_2})\n
    &=-\left(\mathrm{Re}(a^\dagger_{j_1}a_{k_1})+i~\mathrm{Im}(a^\dagger_{j_1}a_{k_1})\right)\nn\times
    \left(\mathrm{Re}(a^\dagger_{j_2}a_{k_2})+i~\mathrm{Im}(a^\dagger_{j_2}a_{k_2})\right)\n
    &=-\mathrm{Re}(a^\dagger_{j_1}a_{k_1})\mathrm{Re}(a^\dagger_{j_2}a_{k_2})\nn
    -i\mathrm{Re}(a^\dagger_{j_1}a_{k_1})\mathrm{Im}(a^\dagger_{j_2}a_{k_2})\nn
    -i\mathrm{Im}(a^\dagger_{j_1}a_{k_1})\mathrm{Re}(a^\dagger_{j_2}a_{k_2})\nn
    +\mathrm{Im}(a^\dagger_{j_1}a_{k_1})\mathrm{Im}(a^\dagger_{j_2}a_{k_2})\label{eqn:decomp1}
\end{align}
\begin{align}
    \langle a^\dagger_{j_1}a^\dagger_{j_2}a_{k_1}a_{k_2}\rangle&=
    -\left\langle\mathrm{Re}(a^\dagger_{j_1}a_{k_1})\mathrm{Re}(a^\dagger_{j_2}a_{k_2})\right\rangle\nn
    +\left\langle\mathrm{Im}(a^\dagger_{j_1}a_{k_1})\mathrm{Im}(a^\dagger_{j_2}a_{k_2})\right\rangle\label{eqn:decomp2}
\end{align}
where the final step discards imaginary components that are expected to vanish in expectation for real-valued spin-orbitals.
Measurement of each term in the last line requires measuring the real and imaginary components (of different operators) simultaneously and multiplying their eigenvalues before averaging, in the same way that Pauli string expectation values can be computed from individual Pauli measurements.
The number of components in the sum grows as $2^p$ for an order $p$ operator. However, since the maximum order is fixed, the decomposition contributes only a constant factor to the required number of measurements.

Since this method utilises the same circuits as the uncorrelated method it can be made to allow post-selection based on both the total number and total spin-projection operators.
To ensure the latter is possible it is necessary to note that there are multiple ways to pair the indices in the first step of the decomposition.
By pairing indices with the same spin (if this is not possible then the excitation must have vanishing expectation value by the arguments in~\ref{app:filtering}), the total spin-projection symmetry is preserved and post-selection is possible.
The decomposition can be generalised to cases where there are non-unique indices by first factoring out all non-unique indices and measuring these logical qubits in the computational basis.

\subsection{Grouping of mutually commuting excitations}\label{app:grouping}
Since there are 940 non-trivial excitations to measure each of which requires up to 8 different measurement bases, it is desirable to group mutually commuting excitations to be measured simultaneously.
This grouping is performed by multiple applications of a greedy algorithm, a brief outline of which is given here, with an example given in~\ref{app:example}.

First we define a representation of a measurement basis as a list of qubit interactions. Each qubit can interact with one other qubit (or no other qubit for a number measurement $a^\dagger_ja_j$).
For example, a measurement basis might be written as: $(0,0),(1,2)$, indicating a number measurment on qubit 0 and a joint measurement on qubits 1 and 2.
A basis measures an excitation if each creation index in the excitation is interacted with an annihilation index. Note that at this stage no distinction is made between the two different interactions given in Figure~\ref{fig:G2}.

The first stage of the algorithm partitions excitations into commuting (though not yet measurable) bases.
The second stage partitions each basis further to allow measurement and can be mapped to the more well-known tensor-product-basis grouping problem.

Each partitioning is performed by iterating over the operators to be measured, selecting the `best' basis and updating it. If no compatible bases are found a new basis is appended to the set.
Using this method the 940 excitations (with multiple components each) are partitioned into 200-204 measurement bases.
Due to the greedy algorithm, the final number of bases is dependent on the order in which the excitations are presented to the algorithm and since the different trial-states result in different arrangements of spin-orbitals the ordering of excitations and hence number of bases is different between the two states.

\subsection{Routing of logical qubits}
\includefigrouting{t}
An additional difficulty that has not been considered in the previous section is that of routing each pair of qubits together with as little circuit depth increase as possible.
It is possible to show that each qubit can be interacted with each other qubit in linear depth~\cite{Kivlichan_LinearDepth_2018}, while the problem here is for each qubit to be interacted with a \emph{specific} partner.
Therefore the additional circuit depth should grow at most linearly with the number of qubits but naively routing qubit pairs sequentially often generates prohibitively deep circuit extensions.
Additionally it is not hard to design an example for which the brickwork circuit~\cite{Kivlichan_LinearDepth_2018} also performs poorly compared to solutions generated `by inspection'.
To solve this systematically we convert the problem to an integer linear program~\cite{Mooney_Routing_2022} and solve using Gurobi~\cite{Gurobi_2023}.
The binary variables we optimise are $x_{jk,t}^{(c)}$ and $y_{jk,t}^{(c)}$ where the $x$-variables represent a logical qubit from pair $c$ moving from physical qubit $j$ to an adjacent (or the same) qubit $k$ at time $t$ and the $y$ variables represent a logical qubit from pair $c$ on physical qubit $j$ interacting with its partner on adjacent qubit $k$ at time $t$.
Two constraints on the $y$ variables immediately become apparent
\begin{itemize}
\item $y_{jk,t}^{(c)}=y_{kj,t}^{(c)}$, i.e. interactions must be between two logical qubits from the same pair, in the same timestep
\item $y_{jj,t}^{(c)}=0$, i.e. two interacting logical qubits cannot be on the same physical qubit
\end{itemize}
In addition we use the following constraints:
\begin{enumerate}
\item Enforce starting positions of logical qubits ($S_j^{(c)}$ is 1 if a logical qubit from pair $c$ starts on qubit $j$ and 0 otherwise),
\begin{align}
    \sum_k\left[x_{jk,0}^{(c)}+y_{jk,0}^{(c)}\right]=S_j^{(c)},
\end{align}
\item The number of logical qubits entering (or remaining on) a physical qubit at a given timestep must equal the number of logical qubits leaving (or remaining on) that physical qubit in the next timestep,
\begin{align}
    \sum_k\left[x_{jk,t}^{(c)}-x_{kj,t+1}^{(c)}+y_{jk,t}^{(c)}-y_{kj,t+1}^{(c)}\right]=0,
\end{align}
\item The number of logical qubits entering a physical qubit at a given timestep must not be greater than~1,
\begin{align}
    \sum_{kc}\left[x_{jk,t}^{(c)}+y_{jk,t}^{(c)}\right]\le1,
\end{align}
\item If a logical qubit moves from physical qubit $j$ to $k$ at timestep $t$, then any logical qubit on $k$ at $t$ must move to $j$ (i.e. enforce swapping behaviour),
\begin{align}
    \sum_c\left[x_{jk,t}^{(c)}+\sum_{j'\ne j}x_{kj',t}^{(c)}\right]\le 1,
\end{align}
\item Interactions must occur exactly once for each pair,
\begin{align}
    \sum_{jkt}\left[y_{jk,t}^{(c)}\right]=2.
\end{align}
\end{enumerate}
Note that the conditions on $y$ can be used to analytically simplify the implementation of the conditions above but have not been substituted here to maintain readability.

Example outputs from the qubit routing optimisation are shown in Figure~\ref{fig:routing}.

\section{Noise mitigation}\label{app:mitigation}
In the near-term fault tolerant quantum computation (i.e. error correction) is not possible due to hardware constraints. Instead, various schemes have been proposed to mitigate hardware noise, usually at the cost of running additional circuits. The methods used in this work are detailed below.
\subsection{Quantum readout-noise mitigation}\label{app:QREM}
One potential source of errors occurs during the qubit-readout process, where the qubit state may be misclassified.
Quantum Readout-Error Mitigation (QREM,~\cite{Maciejewski_QREM_2020}) aims to address these readout errors by calibrating the readout procedure on known computational basis states.
By creating and measuring a selection of computational basis states, the assignment matrix of each qubit can be formed.
This matrix represents the transformation from the ideal output probability vector to the noisy vector.
By inverting this matrix the transformation from noisy to ideal vector can be obtained.
In the implementation used here~\cite{Nation_M3_2021}, readout error on qubit $j$ is averaged over (a subset of) the states of the other qubits but is not explicitly treated as being correlated noise.
Therefore the assignment matrices are $2\times2$ matrices and are efficiently invertible.

In practice QREM may lead to non-physical quasi-probabilities.
To enforce physicality the `negative probability' can be redistributed evenly between measurement outcomes with positive probability~\cite{Michelot_PhysicalProbabilities_1986} (see also~\cite{Smolin_PhysicalEigenvalues_2012}).

\subsection{Symmetry verification}\label{app:SV}
Due to the physical motivation for the UCC-based trial-state, the electron number and total electronic spin are fixed for any value of the trial-state parameters.
Additionally, since we are able to measure both of these quantities simultaneously with the RDM elements through the fermionic measurement procedure described in~\ref{app:RDM}, they can be used for symmetry verification~\cite{BonetMonroig_SV_2018}.
i.e. each time a bitstring is measured, if the corresponding electron number or total electronic spin is not the value expected from the trial-state design, then we can say with certainty that an error has occurred and discard the result.
In this way only errors that commute with both symmetries are not detected and are able to influence the resulting energy.

Such a post-selection strategy necessarily leads to a reduction in the effective number of counts.
In the experiments performed here, only 56\% and 22\% of measurement shots are accepted for the 4- and 6-excitation trial-circuits respectively. 
Note that generating bitstrings at random would have a 14\% acceptance rate.

\subsection{Reduced density matrix rescaling}\label{app:rescaling}
The reduced density matrices as given by Equation~\ref{eqn:rdm} are only well-defined for states with a fixed number of electrons, otherwise the use of Equation~\ref{eqn:reconstruction} can lead to rapid accumulation of error.
Unfortunately, and somewhat counter-intuitively, symmetry verification alone is not sufficient to enforce the correct electron number for the RDM.
This is because symmetry verification enforces the correct correlations only between elements of the same measurement basis, so properties that are reconstructed from measurements in multiple bases can violate symmetry constraints.
To resolve this, it can be noted that the electron number can be directly related to the trace of the ideal $p$-RDM by
\begin{align}
    \mathrm{Tr}({_\mathrm{ideal}R})&\equiv\sum_{j_1<j_2\dots<j_p}{_\mathrm{ideal}R}_{j_1j_2\dots j_p}^{j_1j_2\dots j_p}\n
    &=\frac{N_e!}{p!(N_e-p)!},
\end{align}
and to enforce the correct trace (and therefore electron number) it is possible to simply rescale the RDM by the ratio of the ideal trace to the experimental trace~\cite{Tilly_QRDM_2021}, i.e.
\begin{align}
    _\mathrm{correted}R=\frac{N_e!}{p!(N_e-p)!}\frac{1}{\mathrm{Tr}({_\mathrm{noisy}R})}\cdot{_\mathrm{noisy}R}.
\end{align}

We note that the correct trace is a necessary, but not sufficient, condition for the $p$-RDM to represent a valid physical state. In general, the RDM needs to satisfy the $N$-representability conditions~\cite{
Coleman_RDMs_1963,
Knight_RDMs_2022,
Mazziotti_NRepresentability_2016,
Rubin_NRepresentability_2018,
Cai_QuantumErrorMitigation_2023}. 

Using $\mathbf{j}$ to represent a list of indices, $j_1j_2\dots$, for brevity, then a (still insufficient) list of necessary conditions for validity of the $p$-RDM is~\cite{Rubin_NRepresentability_2018}:
\begin{enumerate}
    \item Hermiticity:
    \begin{align}
        R_\mathbf{j}^\mathbf{k}=(R_\mathbf{k}^\mathbf{j})^*,
    \end{align}
    \item Antisymmetry:
    \begin{align}
        R_{P(\mathbf{j})}^\mathbf{k}=\sigma_PR_\mathbf{k}^\mathbf{j},\n
        R_\mathbf{j}^{P(\mathbf{k})}=\sigma_PR_\mathbf{k}^\mathbf{j},
    \end{align}
    where $P(\mathbf{j})$ is a permutation of the indices contained in $\mathbf{j}$ and $\sigma_P=\pm1$ is the sign of the permutation.
    \item Contraction:
    \begin{align}
        R_{j_1\dots j_p}^{k_1\dots k_p}&=\frac{1}{N_e-p}\sum_{l}R_{j_1\dots j_pl}^{k_1\dots k_pl},
    \end{align}
    \item Trace:
    \begin{align}
        \mathrm{Tr}(R)=\frac{N_e!}{p!(N_e-p)!},
    \end{align}
    \item Positive-semidefiniteness:
    \begin{align}
        R\succeq0.
    \end{align}
\end{enumerate}
Further constraints can be derived from (for example) the requirement that the trial-state is an eigenstate of the spin-projection operator~\cite{Rubin_NRepresentability_2018,Knight_RDMs_2022}. Note that in this work, the Hermiticity and antisymmetry conditions are used to reduce the number of required measurements in~\ref{app:filtering}, the contraction condition is exactly Equation~\ref{eqn:reconstruction} which is used to extract lower-order expectation values from high order RDMs, and the trace condition is enforced by the rescaling technique.
In future work, additional constraints could potentially be implemented to further detect/mitigate errors as has been proposed~\cite{Rubin_NRepresentability_2018} and performed~\cite{Smart_NRepresentability_2019} for 2-RDMs.

An interesting feature of measuring RDMs is that even with this rescaling technique, if no other error mitigation is applied, the energy estimates it produces have lower accuracy than estimating the energy directly from counts. On the other hand, we see that the variance of the energy estimates is significantly reduced by the RDM method. This is advantageous because many error mitigation techniques improve accuracy at the cost of an increased variance. In combination (in particular the noise calibration described below), the RDM method results in both more accurate and more precise (lower variance) estimates than calculating energy estimates directly from counts.

\subsection{Reference-state noise calibration}\label{app:calibration}
\includefigcalibration{t}
Another powerful error mitigation technique is based on the existence of reference circuits~\cite{LolurSkogh_ReferenceStates_2023,Czarnik_CDR_2021}, circuits with similar overall structure to the trial-state but with a limited number of non-Clifford gates.
The simplest reference-state for the circuit used here is the Hartree-Fock state, which can be prepared by setting all the variational parameters to 0.
Since properties of the Hartree-Fock state can be efficiently evaluated classically, comparison between classical and quantum computations can be used to `learn' the noise model of the device.
Here, a global white noise model is assumed,
\begin{align}
    \rho\rightarrow\rho^\mathrm{noisy}&=(1-q)\rho+qJ,\n
    \Rightarrow\mathrm{Tr}(\mathcal{H}\rho^\mathrm{noisy})&=(1-q)\mathrm{Tr}(\mathcal{H}\rho)+q\mathrm{Tr}(\mathcal{H}J),\label{eqn:whitenoise}
\end{align}
where $q$ represents the effective global error rate and $J$ is related to the maximally mixed state (concretely it is the image of the maximally mixed state after application of other error mitigation techniques e.g. post-selection will first zero-out some entries of the mixed-state, $I/2^{N_s}$, then renormalise).
This model has been seen to be reasonable for certain classes of circuits~\cite{Dalzell_WhiteNoise_2021} and circuits could be made to better obey the model through randomised compiling techniques~\cite{Hashim_RandomisedCompiling_2021}.
Using the classical (noiseless) and quantum (noisy) expectation values from the Hartree-Fock state, an estimate, $\hat{q}$, for the error rate can be obtained by solving Equation~\ref{eqn:whitenoise}.
In practice, we find that the effective noise level is around 8\% for values reconstructed from counts and 22\% for values reconstructed from RDMs, which are in turn reconstructed from counts (however, the RDM method is seen to greatly reduce the variance in the white noise estimate which ultimately allows for a more accurate energy estimation).
Following this, quantum computation of properties of the trial-state can be corrected using the noise estimate, $\hat{q}$.
A graphical representation of this process is shown in Figure~\ref{fig:calibration}.

\section{Statistical bootstrapping}\label{app:bootstrapping}
Standard deviations are estimated using a bootstrapping method~\cite{EfronTibshirani_Bootstrapping_1994}. By resampling (classically) $500\times10^5$ times from the probability distributions obtained from the quantum hardware, a distribution of energy estimates is obtained from which error estimates can be extracted.
\end{multicols}

\breakl
\section{Basis grouping example}\label{app:example}
\subsection{Fermionic measurement grouping (first level)}
Consider measuring the spin-conserving excitations for the 2-RDM on 4 spin-orbitals:
\begin{center}
    \begin{tabular}{cccccc}
        $a^\dagger_0a^\dagger_1a_0a_1$\\
        &
        $a^\dagger_0a^\dagger_2a_0a_2$
        &
        $a^\dagger_0a^\dagger_2a_1a_2$
        &
        $a^\dagger_0a^\dagger_2a_0a_3$
        &
        $a^\dagger_0a^\dagger_2a_1a_3$\\
        &&
        $a^\dagger_1a^\dagger_2a_1a_2$
        &
        $a^\dagger_1a^\dagger_2a_0a_3$
        &
        $a^\dagger_1a^\dagger_2a_1a_3$\\
        &&&
        $a^\dagger_0a^\dagger_3a_0a_3$
        &
        $a^\dagger_0a^\dagger_3a_1a_3$\\
        &&&&
        $a^\dagger_1a^\dagger_3a_1a_3$\\
        &&&&&
        $a^\dagger_2a^\dagger_3a_2a_3$
    \end{tabular}
\end{center}
where spin-orbitals 0 and 1 have spin-up and spin-orbitals 2 and 3 have spin-down.
We begin with an empty basis set, $\{\}$, and for each excitation above, search for a commuting basis.
If a basis is found we update it, if no basis is found we add a new one.
\begin{enumerate}
    \item $a^\dagger_0a^\dagger_1a_0a_1$
    \begin{itemize}
        \item $\{\}\rightarrow\{(0,1)\}$
        \item No commuting basis. New basis $(0,1)$.
    \end{itemize}
    \item $a^\dagger_0a^\dagger_2a_0a_2$
    \begin{itemize}
        \item $\{(0,1)\}\rightarrow\left\{\begin{matrix}(0,1)\\(0,0),(2,2)\end{matrix}\right\}$
        \item No commuting basis. New basis $(0,0),(2,2)$. Note that spin-orbitals 0 and 2 have opposite spin, so interacting them is not allowed (fails to commute with the spin-projection operator).
    \end{itemize}
    \item $a^\dagger_0a^\dagger_2a_1a_2$
    \begin{itemize}
        \item $\left\{\begin{matrix}(0,1)\\(0,0),(2,2)\end{matrix}\right\}\rightarrow\left\{\begin{matrix}(0,1),(2,2)\\(0,0),(2,2)\end{matrix}\right\}$
        \item Commuting basis, $(0,1)$ found and updated to $(0,1),(2,2)$.
    \end{itemize}
    \item $a^\dagger_0a^\dagger_2a_0a_3$
    \begin{itemize}
        \item $\left\{\begin{matrix}(0,1),(2,2)\\(0,0),(2,2)\end{matrix}\right\}\rightarrow\left\{\begin{matrix}(0,1),(2,2)\\(0,0),(2,2)\\(0,0),(2,3)\end{matrix}\right\}$
        \item No commuting basis. New basis $(0,0),(2,3)$.
    \end{itemize}
    \item $a^\dagger_0a^\dagger_2a_1a_3$
    \begin{itemize}
        \item $\left\{\begin{matrix}(0,1),(2,2)\\(0,0),(2,2)\\(0,0),(2,3)\end{matrix}\right\}\rightarrow\left\{\begin{matrix}(0,1),(2,2)\\(0,0),(2,2)\\(0,0),(2,3)\\(0,1),(2,3)\end{matrix}\right\}$
        \item No commuting basis. New basis $(0,1),(2,3)$.
    \end{itemize}
    \item $a^\dagger_1a^\dagger_2a_1a_2$
    \begin{itemize}
        \item $\left\{\begin{matrix}(0,1),(2,2)\\(0,0),(2,2)\\(0,0),(2,3)\\(0,1),(2,3)\end{matrix}\right\}\rightarrow\left\{\begin{matrix}(0,1),(2,2)\\(0,0),(1,1),(2,2)\\(0,0),(2,3)\\(0,1),(2,3)\end{matrix}\right\}$
        \item Commuting basis, $(0,0),(2,2)$ found and updated to $(0,0),(1,1),(2,2)$.
    \end{itemize}
    \item $a^\dagger_1a^\dagger_2a_0a_3$
    \begin{itemize}
        \item $\left\{\begin{matrix}(0,1),(2,2)\\(0,0),(1,1),(2,2)\\(0,0),(2,3)\\(0,1),(2,3)\end{matrix}\right\}$
        \item Measuring basis $(0,1),(2,3)$ found.
    \end{itemize}
    \item $a^\dagger_1a^\dagger_2a_1a_3$
    \begin{itemize}
        \item $\left\{\begin{matrix}(0,1),(2,2)\\(0,0),(1,1),(2,2)\\(0,0),(2,3)\\(0,1),(2,3)\end{matrix}\right\}\rightarrow\left\{\begin{matrix}(0,1),(2,2)\\(0,0),(1,1),(2,2)\\(0,0),(1,1),(2,3)\\(0,1),(2,3)\end{matrix}\right\}$
        \item Commuting basis $(0,0),(2,3)$ found and updated to $(0,0),(1,1),(2,3)$.
    \end{itemize}
    \item $a^\dagger_0a^\dagger_3a_0a_3$
    \begin{itemize}
        \item $\left\{\begin{matrix}(0,1),(2,2)\\(0,0),(1,1),(2,2)\\(0,0),(1,1),(2,3)\\(0,1),(2,3)\end{matrix}\right\}\rightarrow\left\{\begin{matrix}(0,1),(2,2)\\(0,0),(1,1),(2,2),(3,3)\\(0,0),(1,1),(2,3)\\(0,1),(2,3)\end{matrix}\right\}$
        \item Commuting basis $(0,0),(1,1),(2,2)$ found and updated to $(0,0),(1,1),(2,2),(3,3)$.
    \end{itemize}
    \item $a^\dagger_0a^\dagger_3a_1a_3$
    \begin{itemize}
        \item $\left\{\begin{matrix}(0,1),(2,2)\\(0,0),(1,1),(2,2),(3,3)\\(0,0),(1,1),(2,3)\\(0,1),(2,3)\end{matrix}\right\}\rightarrow\left\{\begin{matrix}(0,1),(2,2),(3,3)\\(0,0),(1,1),(2,2),(3,3)\\(0,0),(1,1),(2,3)\\(0,1),(2,3)\end{matrix}\right\}$
        \item Commuting basis $(0,1),(2,2)$ found and updated to $(0,1),(2,2),(3,3)$
    \end{itemize}
    \item $a^\dagger_1a^\dagger_3a_1a_3$
    \begin{itemize}
        \item $\left\{\begin{matrix}(0,1),(2,2),(3,3)\\(0,0),(1,1),(2,2),(3,3)\\(0,0),(1,1),(2,3)\\(0,1),(2,3)\end{matrix}\right\}$
        \item Measuring basis $(0,0),(1,1),(2,2),(3,3)$ found.
    \end{itemize}
    \item $a^\dagger_2a^\dagger_3a_2a_3$
    \begin{itemize}
        \item $\left\{\begin{matrix}(0,1),(2,2),(3,3)\\(0,0),(1,1),(2,2),(3,3)\\(0,0),(1,1),(2,3)\\(0,1),(2,3)\end{matrix}\right\}$
        \item Measuring basis $(0,1),(2,2),(3,3)$ found. Note that $(0,0),(1,1),(2,2),(3,3)$ also measures this excitation. At this point the choice between the two is arbitrary.
    \end{itemize}
\end{enumerate}

\subsection{Mapping to Pauli measurement grouping (second level)}
Based on the results for the previous example, we now need to assign the interaction types (i.e. which circuit from Figure~\ref{fig:G2}) to use for each interaction in each basis in the set.
The assignment of interactions in a given basis are independent of the other bases and so we can take the bases in any order.
\begin{enumerate}
    \item $(0,0),(1,1),(2,2),(3,3)$
    \begin{itemize}
        \item Measures $a^\dagger_1a^\dagger_3a_1a_3$, $a^\dagger_0a^\dagger_3a_0a_3$, $a^\dagger_1a^\dagger_2a_1a_2$ and $a^\dagger_0a^\dagger_2a_0a_2$.
        \item Since each qubit is ``interacted'' with itself, the only choice is to measure in the computational basis. This results in measuring the number operators $a^\dagger_ja_j$ (for $j\in[0,3]$) from which the products can be reconstructed. In this case no further partitioning is required.
    \end{itemize}
    \item $(0,1),(2,2),(3,3)$
    \begin{itemize}
        \item Measures $a^\dagger_2a^\dagger_3a_2a_3$, $a^\dagger_0a^\dagger_3a_1a_3$, $a^\dagger_0a^\dagger_2a_1a_2$ and $a^\dagger_0a^\dagger_1a_0a_1$.
        \item Only one of the interactions is between different qubits, therefore this can be mapped to a (trivial) 1 site tensor product basis grouping problem. We will use `site' for these virtual qubits and `qubits' for the actual qubits.
        \begin{itemize}
            \item The interaction $(0,1)$ is mapped to the first (and only) site. We need to decide what basis to measure it in.
            \item The excitation operators are mapped to the Pauli operators to be grouped.
            \begin{itemize}
                \item $a^\dagger_2a^\dagger_3a_2a_3$ doesn't act on the site, so maps to $I$.
                \item $a^\dagger_0a^\dagger_3a_1a_3$ acts on the site as $a^\dagger_0a_1\rightarrow\mathrm{Re}(a^\dagger_0a_1)\rightarrow X$.
                \item $a^\dagger_0a^\dagger_2a_1a_2$ also maps to $X$.
                \item $a^\dagger_0a^\dagger_1a_0a_1$ is a special case and can be extracted from either of the interactions in Figure~\ref{fig:G2} (since they are both number conserving by design) so we can also map this to $I$.
            \end{itemize}
            \item As stated above, the grouping problem in this case is trivial, since there is only 1 site and only one non-identity operator ($X$).
        \end{itemize}
    \end{itemize}
    \item $(0,0),(1,1),(2,3)$
    \begin{itemize}
        \item Measures $a^\dagger_1a^\dagger_2a_1a_3$, $a^\dagger_0a^\dagger_2a_0a_3$
        \item Similar to above, this maps to the problem of 1 site and one operator ($X$).
    \end{itemize}
    \item $(0,1),(2,3)$
    \begin{itemize}
        \item Measures $a^\dagger_1a^\dagger_2a_0a_3$ and $a^\dagger_0a^\dagger_2a_1a_3$
        \item In this case we map to a 2 site tensor product basis problem
        \begin{itemize}
            \item $(0,1)$ and $(2,3)$ become the first and second sites respectively
            \item The excitation operators map as
            \begin{itemize}
                \item $a^\dagger_1a^\dagger_2a_0a_3\rightarrow\mathrm{Re}(a^\dagger_1a_0)\mathrm{Re}(a^\dagger_2a_3)+\mathrm{Im}(a^\dagger_1a_0)\mathrm{Im}(a^\dagger_2a_3)\rightarrow XX+YY$
                \item $a^\dagger_0a^\dagger_2a_1a_3\rightarrow\mathrm{Re}(a^\dagger_0a_1)\mathrm{Re}(a^\dagger_2a_3)+\mathrm{Im}(a^\dagger_0a_1)\mathrm{Im}(a^\dagger_2a_3)\rightarrow XX+YY$
            \end{itemize}
            \item This then leads to the problem of finding a tensor product grouping for $XX$ and $YY$ (which trivially requires 2 bases)
        \end{itemize}
    \end{itemize}
\end{enumerate}
In non-trivial cases i.e. for higher number of spin-orbitals, the Pauli measurement grouping problem is solved using a greedy algorithm analogous to the Fermionic grouping algorithm of the previous section.

\end{document}